\pdfoutput=1  

\documentclass[12pt]{article}

\usepackage{tocloft}
\usepackage{graphicx}
\usepackage{natbib}
\usepackage[protrusion=true,expansion=true]{microtype}
\usepackage{amsmath,amsthm,amsfonts,amssymb,mathrsfs,dsfont,mathtools}
\usepackage{booktabs}
\usepackage{url}
\usepackage{float}
\usepackage{color}
\usepackage{parskip}
\usepackage[explicit]{titlesec}
\usepackage{threeparttable}
\usepackage{setspace}
\usepackage[margin=0.85in]{geometry}
\usepackage[font = onehalfspacing]{caption}
\usepackage[usenames,dvipsnames]{xcolor}
\usepackage[hyperfootnotes=true]{hyperref}
\usepackage[labelfont=bf]{caption}
\usepackage{authblk}
\usepackage{caption}
\usepackage[titletoc,toc,title]{appendix}
\usepackage{enumerate}
\usepackage{soul}
\usepackage{subfig}
\usepackage{subfiles} 
\usepackage[capitalise, nameinlink]{cleveref}
\usepackage{dsfont}
\usepackage{chngpage} 

\newcommand{\E}{\mathbb{E}}

\DeclareMathOperator*{\argmin}{arg\,min}

\usepackage{graphicx}
\newsavebox\CBox

\makeatletter
\newcommand*\bigcdot{\mathpalette\bigcdot@{.5}}
\newcommand*\bigcdot@[2]{\mathbin{\vcenter{\hbox{\scalebox{#2}{$\m@th#1\bullet$}}}}}
\makeatother

\DeclarePairedDelimiter\floor{\lfloor}{\rfloor}

\setstcolor{red} 

\numberwithin{equation}{section}

\hypersetup{colorlinks = true, citecolor = MidnightBlue, linkcolor = MidnightBlue, urlcolor = MidnightBlue}

\titleformat{\section}{\normalfont\large\bfseries}{\thesection}{1em}{#1}
\titleformat{\subsection}{\normalfont\normalsize\bfseries}{\thesubsection}{1em}{#1}
\titleformat{\subsubsection}{\normalfont\normalsize\itshape}{\thesubsubsection}{1em}{#1}
\titlespacing\section{0pt}{12pt plus 4pt minus 2pt}{6pt plus 2pt minus 2pt}
\titlespacing\subsection{0pt}{12pt plus 4pt minus 2pt}{3pt plus 2pt minus 3pt}
\titlespacing\subsubsection{0pt}{12pt plus 4pt minus 2pt}{0pt plus 2pt minus 3pt}

\def\boxit#1{\vbox{\hrule\hbox{\vrule\kern6pt
			\vbox{\kern6pt#1\kern6pt}\kern6pt\vrule}\hrule}}

\definecolor{orange}{rgb}{1,0.5,0}
\definecolor{MyDarkBlue}{rgb}{0,0.08,0.45}

\newtheorem{remark}{Remark}[section]

\newtheorem{assumption}{Assumption}[section]
\newtheorem{Def}{Definition}[section]

\newcommand\blfootnote[1]{%
	\begingroup
	\renewcommand\thefootnote{}\footnote{#1}%
	\addtocounter{footnote}{-1}%
	\endgroup
}

\begin{document}
	
	\title{\Large \bfseries Deep Hedging of Long-Term Financial Derivatives\blfootnote{A GitHub repository with some examples of codes can be found at \href{https://github.com/alexandrecarbonneau}{github.com/alexandrecarbonneau}.} 
	\thanks{
			The author gratefully acknowledges financial support from the Fonds de recherche du Qu\'ebec (FRQNT). He would also like to thank Fr{\'e}d{\'e}ric Godin for his helpful comments and suggestions.}
	}
	
	\author {Alexandre Carbonneau\thanks{{\mbox{\hspace{0.00cm}} \it Email address:} \href{mailto:alexandre.carbonneau@mail.concordia.ca}{alexandre.carbonneau@mail.concordia.ca}}}
	\affil {{\small Concordia University, Department of Mathematics and Statistics, Montr\'eal, Canada}}
	
	\vspace{-10pt}
	\date{ 
		\today}
	
	\maketitle \thispagestyle{empty} 
	
	\vspace{-15pt}
	
	\begin{abstract} 
		\vspace{-5pt}
		This study presents a deep reinforcement learning approach for global hedging of long-term financial derivatives. A similar setup as in \cite{coleman2007robustly} is considered with the risk management of lookback options embedded in guarantees of variable annuities with ratchet features. 
		The deep hedging algorithm of \cite{buehler2019deep} is applied to optimize neural networks representing global hedging policies with both quadratic and non-quadratic penalties. 
		To the best of the author's knowledge, this is the first paper that presents an extensive benchmarking of global policies for long-term contingent claims with the use of various hedging instruments (e.g. underlying and standard options) and with the presence of jump risk for equity. 
		Monte Carlo experiments demonstrate the vast superiority of non-quadratic global hedging as it results simultaneously in downside risk metrics two to three times smaller than best benchmarks and in significant hedging gains. 
		Analyses show that the neural networks are able to effectively adapt their hedging decisions to different penalties and stylized facts of risky asset dynamics only by experiencing simulations of the financial market exhibiting these features.
		%
		Numerical results also indicate that non-quadratic global policies are significantly more geared towards being long equity risk which entails earning the equity risk premium. 
		
		\noindent \textbf{Keywords:} Reinforcement learning; Global hedging; Variable annuity; Lookback option; Jump risk.
		
	\end{abstract} 
	\medskip
	
	\thispagestyle{empty} \vfill \pagebreak
	
	\setcounter{page}{1}
	\pagenumbering{roman}
	
	
	
	\doublespacing
	
	\setcounter{page}{1}
	\pagenumbering{arabic}

	
	\section{Introduction}
	Variable annuities (VAs), also known as segregated funds and equity-linked insurance, are financial products that enable investors to gain exposure to the market through cashflows that depend on equity performance. These products often include financial guarantees to protect investors against downside equity risk with benefits which can be expressed as the payoff of derivatives. For instance, a guaranteed minimum maturity benefit (GMMB) with ratchet feature is analogous to a lookback put option by providing a minimum monetary amount at the maturity of the contract equal to the maximum account value on specific dates (e.g. anniversary dates of the policy). 
	The valuation of VAs guarantees is typically done with classical option pricing theory by computing the expected risk-neutral discounted cashflows of embedded options under an appropriate equivalent martingale measure; see, for instance, \cite{brennan1976pricing}, \cite{boyle1977equilibrium}, \cite{persson1997valuation}, \cite{bacinello2003fair} and \cite{bauer2008universal}. A comprehensive review of pricing segregated funds guarantees literature can be found in \cite{gan2013application}.
	
	During the subprime mortgage financial crisis, many insurers incurred large losses in segregated fund portfolios due in part to poor risk management with some insurers even stopping writing VAs guarantees in certain markets (\cite{zhang2010integrating}). Two categories of risk management approaches are typically used in practice: the actuarial method and the financial engineering method (\cite{boyle1997reserving}). The foremost 
	consist 
	in providing stochastic models for the risk factors and setting a reserve held in risk-free assets to cover the liabilities associated to VAs guarantees with a certain probability (e.g. the Value-at-Risk at $99\%$). The second approach commonly known as \textit{dynamic hedging} entails finding a self-funded sequence of positions in securities to hedge the risk exposure of embedded options. Dynamic hedging is a popular risk management approach among insurance companies and is studied in this current paper; 
	the reader is referred to \cite{hardy2003investment} for a detailed description of the actuarial approach. 

	Financial markets are said to be complete if every contingent claim can be perfectly replicated with some dynamic hedging strategy. In practice, segregated funds embedded options are typically not attainable as a consequence of their many interrelated risks which are very complex to manage such as equity risk, interest rate risk, mortality risk and basis risk. For insurance companies selling VAs with guarantees, market incompleteness entails that some level of residual risk must be accepted as being intrinsic to the embedded options; the identification of optimal hedging policies in such context is thus highly relevant. 
	Nevertheless, the attention of the actuarial literature has predominantly been on the valuation of segregated funds, not on the design of optimal hedging policies. Indeed, the hedging strategies considered are most often suboptimal and are not necessarily in line with the financial objectives of insurance companies.
	One popular hedging approach is the \textit{greek-based policy} where assets positions depend on the sensitivities of the option value (i.e. the value of the guarantee) to different risk factors. \cite{boyle1997reserving} and \cite{hardy2000hedging} delta-hedge GMMBs under market completeness for mortality risk and \cite{augustyniak2017mitigating} delta-rho hedge GMMBs and guaranteed minimum death benefits (GMDBs) in the presence of model uncertainty for both equity and interest rate. 
	An important pitfall of greek-based policies in incomplete markets is their suboptimality by design: they are a by-product of the choice of pricing kernel (i.e. of the equivalent martingale measure) for option valuation, not of an optimization 
	procedure over hedging decisions to minimize residual risk.
	Also, as shown in the seminal work of \cite{harrison1981martingales}, in incomplete markets, there exist an infinite set of equivalent martingale measures each of which is consistent with arbitrage-free pricing and can thus be used to compute the hedging positions (i.e. the greeks). 

	Another strand of literature optimizes hedging policies with local and global criterions. \textit{Local risk minimization} (\cite{follmer1988hedging} and \cite{schweizer1991option}) 
	consists in choosing assets positions to minimize the periodic risk associated with the hedging portfolio. On the other hand, \textit{global risk minimization} procedures jointly optimize all hedging decisions 
	with the objective of minimizing the expected value of a loss function applied to the terminal hedging error. In spite of their myopic view
	of the hedging problem by not necessarily minimizing the risk associated with hedging shortfalls, local risk minimization procedures are attractive for the risk mitigation of VAs guarantees as they are simple to implement and they 
	have outperformed greek-based hedging in several studies. \cite{coleman2006hedging} and \cite{coleman2007robustly} apply local risk minimization procedures for risk mitigation of GMDBs using standard options with the foremost considering the presence of both interest rate and jump risk and the latter the presence of volatility and jump risk. \cite{kelani2017pricing} extends the work of \cite{coleman2007robustly} in a general L\'evy market by including mortality and transaction costs, and \cite{trottier2018fund} and \cite{trottier2018local} propose a local risk minimization scheme for guarantees in the presence of basis risk.

	Within the realm of total risk minimization, \textit{global quadratic hedging} pioneered by the seminal work of \cite{schweizer1995variance} aims at jointly optimizing all hedging decisions with a quadratic penalty for hedging shortfalls. The latter paper provides a 
	theoretical solution to the optimal policy with a single risky asset (see \cite{remillard2013optimal} for the multidimensional asset case) and 
	\cite{bertsimas2001hedging} develops a tractable solution to the optimal policy relying on stochastic dynamic programming. 
%
	A major drawback of global quadratic hedging is in penalizing equally gains and losses which is naturally not in line with the financial objectives of insurance companies. Alternatively, \textit{non-quadratic global hedging} applies an asymmetric treatment to hedging errors by overly (and most often strictly) penalizing hedging losses. In contrast to global quadratic hedging, there is usually no closed-form solution to the optimal policy, but numerical implementations have been proposed in the literature: \cite{franccois2014optimal} developed a methodology with stochastic dynamic programming algorithms for global hedging with any desired penalty function, \cite{godin2016minimizing} adapts the latter numerical implementation under the Conditional Value-at-Risk measure in the presence of transaction costs and \cite{dupuis2016short} apply global hedging under the semi-mean-square error penalty 
	in the context of short-term hedging for an electricity retailer. The aforementioned studies demonstrated the vast superiority of non-quadratic global hedging over other hedging schemes (e.g. greek-based policies, local risk minimization and global quadratic hedging). Yet, to the best of the author's knowledge, 
	both quadratic and non-quadratic global hedging has seldom 
	been applied for risk mitigation of segregated funds guarantees, or more generally, of long-term contingent claims.\footnote{
	An exception is the work of \cite{ankirchner2014cross} which considers a minimal-variance hedging strategy for VAs guarantees in continuous-time in the presence of basis risk.
	} 
	Moreover, numerical schemes for global hedging are computationally intensive and often rely on solving Bellman's equations which is known to be prone to the curse of dimensionality \citep{powell2009you}. In the context of dynamically hedging segregated funds guarantees, the latter is a major drawback as it restrains the number of risk factors to consider for the financial market as well as prevents the use of multiple assets in the design of hedging policies.
	A feasible implementation of global hedging for the risk mitigation of VAs guarantees which is flexible to the choice of market features, to the hedging instruments and to the penalty for hedging errors would be desirable.
	
%

	Recently, \cite{buehler2019deep} introduced a deep reinforcement learning (deep RL) algorithm called \textit{deep hedging} to hedge a portfolio of over-the-counter derivatives in the presence of market frictions. 
	The general framework of RL is for an agent to learn over many iterations of an environment how to select sequences of actions to optimize a cost function.
	RL has been applied successfully in many areas of quantitative finance such as algorithmic trading (e.g. \cite{moody2001learning} and \cite{deng2016deep}), portfolio optimization (e.g. \cite{jiang2017deep} and \cite{almahdi2017adaptive}) and option pricing (e.g. \cite{li2009learning}, \cite{becker2019deep} and \cite{carbonneau2020}). Hedging has also received some attention: \cite{halperin2020qlbs} and \cite{kolm2019dynamic} propose TD-learning approaches to the hedging problem
	and \cite{hongkai2020} and \cite{carbonneau2020} deep hedge European options under respectively the quadratic penalty and the Conditional Value-at-Risk measure.
	The deep hedging algorithm trains an agent to learn how to approximate optimal hedging decisions by neural networks
	through many simulations of a synthetic market. This approach is related to the deep learning method of \cite{han2016deep} by directly optimizing policies for stochastic control problems with Monte Carlo simulations. Arguably, the most important benefit of using neural networks to approximate optimal policies is to overcome the curse of dimensionality which arises when the state-space gets too large. 
	
	The contribution of this paper is threefold. First, this study presents a deep reinforcement learning procedure for global hedging long-term financial derivatives which are analogous under assumptions made in this study to embedded options of segregated funds. Our methodological approach which relies on the deep hedging algorithm can be applied 
	for the risk mitigation of any long-term European-type contingent claims (e.g. vanilla, path-dependent) with multiple hedging instruments (e.g. standard options and underlying) under any desired penalty (e.g. quadratic and non-quadratic) and in the presence of different risky assets stylized features (e.g. jump, volatility and regime risk).
	%
%
	%
	The second contribution consists in conducting broad numerical experiments of hedging long-term contingent claims with the optimized global policies. A similar setup as in the work of \cite{coleman2007robustly} is considered with the risk mitigation of ratchet GMMBs
	strictly for financial risks in the presence of jumps for equity. To the best of the author's knowledge, this is the first paper that presents such an extensive benchmarking of quadratic and non-quadratic global policies for long-term options with the use of various hedging instruments and by considering different risky assets dynamics. Such benchmarking would have been inaccessible when relying on more traditional optimization procedures for global hedging such as stochastic dynamic programming due to the curse of dimensionality. Numerical results demonstrate the vast superiority of non-quadratic global hedging as it results simultaneously in downside risk metrics two to three times smaller than best benchmarks and in significant hedging gains. Our results clearly demonstrate that non-quadratic global hedging should be prioritized over other popular dynamic hedging procedures found in the literature as it is tailor-made to match the financial objectives of the hedger by always significantly reducing the downside risk as well as earning large expected positive returns.
	The third contribution is in providing important insights into specific characteristics of the optimized global policies. Monte Carlo experiments indicate that on average, non-quadratic global policies are significantly more bullish than their quadratic counterpart by holding a larger average equity risk exposure which entails earning the equity risk premium.
	Key factors which contribute to this specific characteristic of non-quadratic global policies are identified.
	Furthermore, analyses of numerical results show 
	that the training algorithm is able to effectively adapt hedging policies (i.e. neural networks parameters) to different stylized features of risky asset dynamics only by experiencing simulations of the financial market exhibiting these features.
	
	The paper is structured as follows. \cref{sec:market_description} introduces the notation and the optimal hedging problem. \cref{sec:Methodology} describes the numerical scheme based on deep RL to optimize global hedging policies. \cref{sec:numerical_results} presents benchmarking of the risk mitigation of GMMBs under various market settings. \cref{sec:conclusion} concludes.
	
	
	\section{Hedging of long-term contingent claims}
	\label{sec:market_description}
	This section details the financial market setup and the hedging problem considered in this paper.
	
	\subsection{Market setup}
	The financial market is in discrete-time with a finite time horizon of $T \in \mathbb{N}$ years and $N+1$ known observation dates $\mathcal{T} := \{t_{i} : t_{i}= i \Delta_{N}, i = 0,\ldots, N\}$ 
	with $\Delta_{N} := T/N$. The probability space 
	$(\Omega, \mathcal{F}_{T},\mathbb{P})$ with $\mathbb{P}$ as the physical measure is equipped with the filtration $\mathbb{F} := \{\mathcal{F}_{t_{n}}\}_{n=0}^{N}$ that defines all available information of the financial market to investors. A total of $D+2$ liquid assets are accessible to financial participants with $D+1$ risky assets and one risk-free asset. Let $\{B_{t_{n}}\}_{n=0}^{N}$ be the price process of the risk-free asset where $B_{t_{n}}:= e^{r t_{n}}$ with $r\in\mathbb{R}$ as the annualized continuous risk-free rate. The risky assets include a non-dividend paying stock and $D$ liquid vanilla European-type options such as calls and puts on the stock which expire on observation dates in $\mathcal{T}$. In this context, the specification of two distinct price processes, one at the beginning and one at the end of each trading period, is required. Let $\{\bar{S}_{t_{n}}^{(b)}\}_{n=0}^{N}$ be the risky price process at the beginning of each trading period where $\bar{S}_{t_{n}}^{(b)}:= [S_{t_{n}}^{(0,b)},\ldots,S_{t_{n}}^{(D,b)}]$ are the prices at the beginning of $[t_{n}, t_{n+1})$ with $S_{t_{n}}^{(0,b)}$ and $S_{t_{n}}^{(j,b)}$ respectively as the price of the underlying and of the $j^{\text{th}}$ option. Similarly, let $\{\bar{S}_{t_{n}}^{(e)}\}_{n=0}^{N-1}$ be the risky price process at the end of each trading period where $\bar{S}_{t_{n}}^{(e)}:= [S_{t_{n}}^{(0,e)},\ldots,S_{t_{n}}^{(D,e)}]$ are the prices at the end of $[t_{n}, t_{n+1})$ before the next rebalancing at $t_{n+1}$. For the tradable options, if the $j^{\text{th}}$ option matures at $t_{n+1}$, then $S_{t_{n}}^{(j,e)}$ is the payoff of the derivative and $S_{t_{n+1}}^{(j,b)}$ is the price of a new contract with the same characteristics (i.e. same payoff function and time-to-maturity). For the underlying, the equality $S_{t_{n+1}}^{(0,b)} = S_{t_{n}}^{(0,e)}$ holds $\mathbb{P}$-a.s. for $n=0,\ldots,N-1$. 

	
	This paper studies the problem of hedging long-term contingent claims embedded in segregated funds guarantees by means of dynamic hedging with a similar setup as in the work of \cite{coleman2007robustly}.
	While the latter paper considers the presence of both jump risk and volatility risk for the equity, the current work strictly assesses the impact of jump risk on the risk management of long-term contingent claims. We note that the methodological approach presented in \cref{sec:Methodology} for 
	optimizing
	global policies can easily be adapted to the presence of additional risk factors for equity (e.g. volatility risk and regime risk). 
	%
	For the rest of the paper, assume that mortality risk can be completely diversified away and let $T$ be the known maturity in years of the embedded guarantee to be hedged. This assumption can be motivated by the fact that in practice, insurance companies can significantly reduce the impact of mortality risk on their segregated funds portfolios by insuring additional policies. Furthermore, all VAs are assumed to be held until expiration (i.e. no lapse risk) and their values are linked to a liquid index such as the S\&P500 which implies no basis risk. 
	
	In this study, the option embedded in VAs is a GMMB with an annual ratchet feature which provides a payoff at time $T$ of the maximum anniversary account value. The anniversary dates of the equity-linked insurance account are assumed to form a subset of the observation dates, i.e. $\{0,1,\ldots,T\} \subseteq \mathcal{T}$.
	Let $\{Z_{t_{n}}\}_{n=0}^{N}$ be the running maximum anniversary value process of the equity-linked account\footnote{
		$\floor{\cdot}: \mathbb{R} \rightarrow \mathbb{R}$ is the floor function, i.e. $\floor{x}$ is the largest integer smaller or equal to $x$.
	}:
	\begin{equation*}
	Z_{t_{n}} = 
	\begin{cases}
	\max(S_{0}^{(0,b)},\ldots,S_{m}^{(0,b)}), &\mbox{if } \floor*{t_{n}} = m \mbox{ and } m \in \{0,\ldots,T-1\},\\
	\max(S_{0}^{(0,b)},\ldots,S_{T-1}^{(0,b)}), &\mbox{if } t_n = T.\\
	\end{cases}
	\end{equation*}
%
	The payoff of the GMMB with annual ratchet can be expressed as the account value at time $T$ plus a lookback put option payoff
	\begin{align}
	\max(S_{0}^{(0,b)},\ldots,S_{T}^{(0,b)}) &= \max(\max(S_{0}^{(0,b)},\ldots,S_{T-1}^{(0,b)}), S_{T}^{(0,b)}) \nonumber
	\\ &= \max(Z_{T}-S_{T}^{(0,b)},0) + S_{T}^{(0,b)}. \label{eq:ref_payoff_ratchet}
	\end{align}
	Thus, the assumptions of market completeness with respect to mortality risk and lapse risk considered in this paper entail that the risk exposure of the insurer selling a GMMB\footnote{
		\cite{coleman2007robustly} consider the problem of hedging a ratchet GMDB with a fixed and known maturity $T$. The use of a fixed maturity in the latter paper is motivated by assuming market completeness under mortality risk and hedging the expected loss of the guarantee. While the current paper considers the risk mitigation of a GMMB instead of a GMDB, assumptions made in both papers (i.e. no mortality risk and lapse risk) entail that the benefits of the two guarantees are equivalent and result in the same lookback put option to hedge as in \eqref{eq:ref_payoff_lookback_option}. 
	} is equivalent to holding short position in a long-term lookback option of fixed maturity $T$ and of payoff $\Phi : \mathbb{R} \times \mathbb{R}^{T} \rightarrow [0,\infty)$:
	\begin{align}
	\Phi(S_{T}^{(0,b)},Z_{T}):=\max(Z_{T}-S_{T}^{(0,b)}, 0). \label{eq:ref_payoff_lookback_option}
	\end{align} 
	Let $\delta := \{\delta_{t_{n}}\}_{n=0}^{N}$ be a trading strategy used by the hedger to minimize his risk exposure to $\Phi$ where for $n=1,\ldots,N$, $\delta_{t_{n}} := (\delta_{t_{n}}^{(0)}, \ldots, \delta_{t_{n}}^{(D)}, \delta_{t_{n}}^{(B)})$ is a vector containing the number of shares held in each asset during the period $(t_{n-1},t_{n}]$ with $\delta_{t_{n}}^{(0:D)} :=(\delta_{t_{n}}^{(0)}, \ldots, \delta_{t_{n}}^{(D)})$ and $\delta_{t_{n}}^{(B)}$ respectively as the positions in the $D+1$ risky assets and in the risk-free asset. The initial portfolio (at time $0$ before the first trade) is invested strictly in the risk-free asset. Also, for convenience, all options used as hedging instruments have one period maturity, i.e. they are traded once and held until expiration. 
	Here is an additional assumption considered for the rest of the paper.
	\begin{assumption}
	\label{assump_2}
		The market is liquid and trading in risky assets does not affect their prices. 
	\end{assumption}
	Before describing the optimization problem of hedging $\Phi$, some well-known concepts in the mathematical finance literature must be described. The reader is referred to \cite{lamberton2011introduction} for additional details. Let $\{G_{t_{n}}^{\delta}\}_{n=0}^{N}$ be the discounted gain process associated with the strategy $\delta$ where $G_{t_{n}}^{\delta}$ is the discounted gain at time $t_{n}$ prior to rebalancing. $G_{0}^{\delta} := 0$ and 
	\begin{align}
	G_{t_{n}}^{\delta}:= \sum_{k=1}^{n} \delta_{t_{k}}^{(0:D)} \bigcdot (B_{t_{k}}^{-1} \bar{S}_{t_{k-1}}^{(e)} - B_{t_{k-1}}^{-1} \bar{S}_{t_{k-1}}^{(b)}), \quad n = 1,2,\ldots,N, \label{eq:ref_disc_gain_process_def}
	\end{align}
	where $\bigcdot$ is the dot product operator.\footnote{
		If $X = [X_{1},\ldots,X_{K}]$ and $Y = [Y_{1},\ldots,Y_{K}]$, $X \bigcdot Y := \sum_{i=1}^{K}X_{i}Y_{i}$.
	} Moreover, let $\{V_{t_{n}}^{\delta}\}_{n=0}^{N}$ be hedging portfolio values for a trading strategy $\delta$ where $V_{t_{n}}^{\delta}$ is the value prior to rebalancing at time $t_{n}$:
	\begin{align}
	V_{t_{n}}^{\delta}:=\delta_{t_{n}}^{(0:D)} \bigcdot \bar{S}_{t_{n-1}}^{(e)} + \delta_{t_{n}}^{(B)}B_{t_{n}}, \quad n=1,\ldots,N, \label{eq_ref:portfolio_value_func}
	\end{align}
	and $V_{0}^{\delta}:=\delta_{0}^{(B)}$ since the initial capital amount is assumed to be strictly invested in the risk-free asset. In this paper, the trading strategies considered require no cash infusion nor withdrawal except at the initialization of the contract (i.e. at time $0$). Such strategies are called \textit{self-financing}. More precisely, the hedging strategy $\delta$ is said to be self-financing if it is predictable\footnote{
	$X = \{X_{n}\}_{n=0}^{N}$ with $X_{n} = [X_{n}^{(1)},\ldots,X_{n}^{(K)}]$ is $\mathbb{F}$-predictable if for $j = 1,\ldots,K,$ $X_{0}^{(j)} \in \mathcal{F}_{0}$ and $X_{n+1}^{(j)} \in \mathcal{F}_{n}$ for $n = 0,\ldots,N-1$. 
	} and if
	\begin{align}
	\delta_{t_{n+1}}^{(0:D)} \bigcdot \bar{S}_{t_{n}}^{(b)} + \delta_{t_{n+1}}^{(B)}B_{t_{n}} = V_{t_{n}}^{\delta}, \quad n = 0,1,\ldots,N-1. \label{eq:ref_self_financing}
	\end{align}
	Lastly, let $\Pi$ be the set of admissible trading strategies for the hedger which consists of all sufficiently well-behaved self-financing strategies.
	\begin{remark}
		It can be shown that $\delta$ is self-financing if and only if $V_{t_{n}}^{\delta} = B_{t_{n}}(V_{0}^{\delta} + G_{t_{n}}^{\delta})$ for $n = 0,1,\ldots,N.$ See for instance \cite{lamberton2011introduction}.
	\end{remark}

	
	\subsection{Optimal hedging problem}
	\label{subsec:optimal_hedging}
	The optimization problem of hedging the risk exposure associated to a short position in the long-term lookback option is now formally defined. For the hedger, the problem consists in the design of a trading policy which minimizes a \textit{penalty}, also referred to as a \textit{loss function}, of the difference between the payoff of the lookback option and the hedging portfolio value at maturity (i.e. the \textit{hedging error} or \textit{hedging shortfall}). Strategies embedded in such policies are called \textit{global hedging strategies} as they are jointly optimized over all hedging decisions until the maturity of the lookback option. 
	Let $\mathcal{L} : \mathbb{R} \rightarrow \mathbb{R}$ be a loss function for the hedging error. For the rest of the paper, assume without loss of generality that the position in the hedging portfolio is long, and that all assets and penalties are well-behaved and integrable enough. Specific conditions are beyond the scope of this study.

	\begin{Def}{(\textit{Global risk exposure})}
		\label{Def:risk_exposure_short}
		 Define $\epsilon(V_0)$ as the global risk exposure of the short position in $\Phi$ under optimal hedge if the value of the initial hedging portfolio is $V_{0} \in \mathbb{R}$:
		\begin{align}
		\epsilon(V_0) &:= \underset{\delta \in \Pi}{\min} \, \E\left[\mathcal{L}\left(\Phi(S_{T}^{(0,b)},Z_{T}) - V_{T}^{\delta}\right)  \right], \label{eq:risk_short}
		\end{align}
		where the expectation is taken with respect to the physical measure. 
	\end{Def}

	
	\begin{remark}
		An assumption implicit to \cref{Def:risk_exposure_short} is that the minimum \eqref{eq:risk_short} is indeed attained by some trading strategy, i.e. that the infimum is in fact a minimum. The identification of conditions which ensure that this assumption is satisfied are left out-of-scope.
	\end{remark}

	The following defines the optimal hedging strategy for $\Phi$ given the initial capital investment and the loss function for hedging errors.
	\begin{Def}{(\textit{Optimal hedging strategy)}} 
		Let $\delta^{\star}(V_{0})$ be the optimal hedging strategy corresponding to the global risk exposure of the hedger if the initial portfolio value is $V_{0} \in \mathbb{R}$:
		\begin{align}
		\delta^{\star}(V_0) &:= \underset{\delta \in \Pi}{\argmin} \, \E\left[\mathcal{L}\left(\Phi(S_{T}^{(0,b)},Z_{T}) - V_{T}^{\delta}\right)\right]. \label{eq:optimal_hedge_short}
		\end{align}
	\end{Def}

	In a realistic setting, the choice of loss function should reflect the financial objectives and the risk aversion of the hedger. One example of penalty which has been extensively studied in the hedging literature is the mean-square error (MSE): $\mathcal{L}(x) = x^{2}$. This penalty entails that hedging gains and losses are treated equally which could be desirable for a financial participant who has to provide a price quote on a security prior to knowing his position (long or short). In the context of this paper where the position in $\Phi$ is always short, penalizing hedging gains is clearly undesirable for the hedger. The corresponding loss function to the MSE that penalizes only hedging losses is the semi-mean-square error (SMSE): $\mathcal{L}(x) = x^{2}\mathds{1}_{\{x > 0\}}$. 
	While the MSE and SMSE are the only penalties considered in numerical experiments of \cref{sec:numerical_results}, the optimization procedure for global hedging policies presented in \cref{sec:Methodology} is flexible to any well-behaved penalties (see e.g. \cite{carbonneau2020} for an implementation with the Conditional Value-at-Risk measure).

	The author wants to emphasize that different penalties will often result in different optimal hedging strategies. An extensive numerical study of the impact of the choice of loss function on the hedging policy for the risk management of lookback options is done in \cref{sec:numerical_results}. Moreover, while the numerical section of this paper strictly studies a specific example of long-term option to hedge, namely the lookback option of payoff $\Phi$, the methodological approach to approximate optimal hedging strategies can be applied for any European-type derivative of well-behaved payoff function which can naturally include other VAs guarantees with payoffs analogous to financial derivatives.

	
	\section{Methodology}
	\label{sec:Methodology}
	This section describes the reinforcement learning procedure used to optimize global policies.
	The approach relies on the \textit{deep hedging} algorithm of \cite{buehler2019deep} who showed that a \textit{feedforward neural network} (FFNN) can be used to approximate arbitrarily well optimal hedging strategies in very general financial market conditions. At its core, a FFNN is a parameterized composite function which maps input to output vectors through the composition of a sequence of functions called \textit{hidden layers}. 
	Each hidden layer applies an affine transformation and a nonlinear transformation to input vectors. 
	A FFNN $F_{\theta} : \mathbb{R}^{d_{0}} \rightarrow \mathbb{R}^{\tilde{d}}$ with $L$ hidden layers has the following representation:
	$$F_{\theta}(X):= o \circ h_{L} \circ \ldots \circ h_{1},$$ 
	$$h_{l}(X):=g(W_{l}X + b_{l}), \quad l=1,\ldots,L,$$
	where $W_{l} \in \mathbb{R}^{d_{l} \times d_{l-1}}$ and $b_{l} \in \mathbb{R}^{d_{l} \times 1}$ are respectively known as the weight matrix and bias vector of the $l^{\text{th}}$ hidden layer $h_{l}$, $g$ is a non-linear function applied to each scalar given as input and $o : \mathbb{R}^{d_{L}} \rightarrow \mathbb{R}^{\tilde{d}}$ is the output function which applies an affine transformation to the output of the last hidden layer $h_{L}$ and possibly also a nonlinear transformation with the same range as $F_{\theta}$. Furthermore, the \textit{trainable parameters} $\theta$ is the set of all weight matrices and bias vectors which are \textit{learned} (i.e. fitted in statistical terms) by minimizing a specified cost function.
	 
	In the current study, the type of neural network considered for functions representing hedging policies is from the family of \textit{recurrent neural networks} (RNNs, \cite{rumelhart1986learning}), a class of neural networks which maps input sequences to output sequences. The architecture of RNNs is similar to FFNNs but differs by having self-connections in hidden layers: each hidden layer is a function of both an input vector from the current time-step and an output vector from the hidden layer of the previous time-step, hence the name \textit{recurrent}.  
	More formally, for an input vector $X_{t_{n}}$ at time $t_{n}$, the time-$t_{n}$ output of the hidden layer is computed as $h_{t_{n}} = f(h_{t_{n-1}}, X_{t_{n}})$ for some time-independent function $f$.\footnote{
	Here, $h_{t_{n-1}}$ and $h_{t_{n}}$ are to be understood for convenience as output vectors from hidden layers and not as mappings.
	} 
	In contrast to FFNNs, feedback loops in hidden layers entail that each output is dependent of past inputs which makes RNNs more appropriate for time-series modeling.
	The type of RNN considered for dynamic hedging in this study is the \textit{long short-term memory} (LSTM) introduced by \cite{hochreiter1997long}. This choice of neural network is motivated by recent results of \cite{buehler2019deep_2} who showed that LSTMs hedging policies
	are more effective for the risk mitigation of path-dependent contingent claims than FFNNs policies.
	Additional remarks are made in subsequent sections to motivate the choice of an LSTM for the specific setup considered in the current paper. For more general information about RNNs, the reader is referred to Chapter $10$ of \cite{goodfellow2016deep} and the many references therein.

	The LSTM architecture is now formally defined. The application of LSTMs as functions representing global hedging policies is described in \cref{subsec_LSTM_def_hedging}.
	In what follows, the time-steps are the same as the observation dates of the financial market. 
	%
	\begin{Def}{(LSTM)}
	\label{def:LSTM}
	Let $F_{\theta}:\mathbb{R}^{N} \times \mathbb{R}^{d_{\text{in}}} \rightarrow \mathbb{R}^{N} \times \mathbb{R}^{d_{\text{out}}}$ be an LSTM which maps the sequence of feature vectors $\{X_{t_{n}}\}_{n=0}^{N-1}$ to $\{Y_{t_{n}}\}_{n=0}^{N-1}$ where $X_{t_{n}}$ and $Y_{t_{n}}$ are respectively two vectors of dimensions $d_{\text{in}}, d_{\text{out}}\in \mathbb{N}$. Let $\text{sigm}(\cdot)$ and $\text{tanh}(\cdot)$ be the sigmoid and hyperbolic tangent functions applied element-wise to each scalar given as input.\footnote{
		For $X:=[X_1,\ldots,X_K]$, $\text{sigm}(X) := \left[\frac{1}{1 + e^{-X_1}}, \ldots, \frac{1}{1 + e^{-X_K}}\right]$ and $\text{tanh}(X) := \left[\frac{e^{X_1}-e^{-X_1}}{e^{X_1}+e^{-X_1}}, \ldots, \frac{e^{X_K}-e^{-X_K}}{e^{X_K}+e^{-X_K}}\right]$.
	} For $H \in \mathbb{N}$, the computation of $F_{\theta}$ at each time-step consists of $H$ LSTM cells which are analogous to but more complex than RNNs hidden layers. Each LSTM cell outputs a vector of $d_{j}$ neurons denoted as $h_{t_{n}}^{(j)} \in \mathbb{R}^{d_{j} \times 1}$ at time $t_{n}$ for $d_j \in \mathbb{N}$ and $j=1,\ldots,H$. 
	More precisely, the computation done by the $j^{\text{th}}$ LSTM cell at time $t_{n}$ is as follows\footnote{
	At time $0$ (i.e. $n=0$), the computation of the $H$ LSTM cells is the same as in \eqref{eq:ref_LSTM_cell} with $h_{t_{-1}}^{(j)}$ and $c_{t_{-1}}^{(j)}$ as vectors of zeros of dimensions $d_{j}$ for $j=1,\ldots,H$.
	%
	}:
%
%
	\begin{align}
	i_{t_{n}}^{(j)} &= \text{sigm}(W_i^{(j)}[h_{t_{n-1}}^{(j)}, h_{t_{n}}^{(j-1)}] + b_i^{(j)}), \nonumber
	\\ f_{t_{n}}^{(j)} &= \text{sigm}(W_f^{(j)}[h_{t_{n-1}}^{(j)}, h_{t_{n}}^{(j-1)}] + b_f^{(j)}), \nonumber
	\\ o_{t_{n}}^{(j)} &= \text{sigm}(W_o^{(j)}[h_{t_{n-1}}^{(j)}, h_{t_{n}}^{(j-1)}] + b_o^{(j)}), \nonumber
	%
	%
	\\ c_{t_{n}}^{(j)} &= f_{t_{n}}^{(j)} \circ c_{t_{n-1}}^{(j)} + i_{t_{n}}^{(j)} \circ \text{tanh}(W_c^{(j)}[h_{t_{n-1}}^{(j)}, h_{t_{n}}^{(j-1)}] + b_c^{(j)}), \nonumber
	\\ h_{t_{n}}^{(j)} &= o_{t_{n}}^{(j)} \circ \text{tanh}(c_{t_{n}}^{(j)}), \label{eq:ref_LSTM_cell}
	%
	\end{align}
	where $[\cdot\, ,\cdot]$ and $\circ$ denote respectively the concatenation of two vectors and the Hadamard product (i.e. the element-wise product) and
	\begin{itemize}
		\item $W_i^{(1)}, W_f^{(1)}, W_o^{(1)}, W_c^{(1)} \in \mathbb{R}^{d_1  \times (d_1 + d_{\text{in}})}$ and $b_i^{(1)}, b_f^{(1)}, b_o^{(1)}, b_c^{(1)} \in \mathbb{R}^{d_1 \times 1}$. 
		\item If $H \geq 2$: $W_i^{(j)}, W_f^{(j)}, W_o^{(j)}, W_c^{(j)} \in \mathbb{R}^{d_j  \times (d_{j} + d_{j-1})}$ and $b_i^{(j)}, b_f^{(j)}, b_o^{(j)}, b_c^{(j)} \in \mathbb{R}^{d_j \times 1}$ for $j=2,\ldots,H$. 
	\end{itemize}
	 At each time-step, the input of the first LSTM cell is the feature vector (i.e. $h_{t_{n}}^{(0)}:=X_{t_{n}}$) and the final output is an affine transformation of the output of the last LSTM cell:
	\begin{align}
	Y_{t_{n}} = W_{y}h_{t_{n}}^{(H)} + b_{y}, \quad n = 0,\ldots,N-1, \label{eq:ref_output_LSTM}
	\end{align}
	where $W_y \in \mathbb{R}^{d_{out} \times d_{H}}$ and $b_y \in \mathbb{R}^{d_{out} \times 1}$.
	Lastly, the set of trainable parameters denoted as $\theta$ consists of all weight matrices and bias vectors:
	\begin{align}
	\theta := \left\{\{W_i^{(j)}, W_f^{(j)}, W_o^{(j)}, W_c^{(j)}, b_i^{(j)}, b_f^{(j)}, b_o^{(j)}, b_c^{(j)}\}_{j=1}^{H}, W_y, b_y\right\}. \label{eq:ref_trainable_params}
	\end{align}
	\end{Def}
	\begin{remark}
		In the deep learning literature, the $i_{t_{n}}^{(j)}$, $f_{t_{n}}^{(j)}$ and $o_{t_{n}}^{(j)}$ are known as input gates, forget gates and output gates. Their architectures have shown to help to alleviate the issue of learning long-term dependencies of time series with classical RNNs as they control the information passed through the LSTM cells. The reader is referred to \cite{bengio1994learning} for more information about this latter pitfall of RNNs and to Chapter $10.10$ of \cite{goodfellow2016deep} and the many references therein for more general information about LSTMs.
	\end{remark}
	
	\subsection{Hedging with an LSTM}
	\label{subsec_LSTM_def_hedging}
	In the context of dynamic hedging, an LSTM maps a sequence of feature vectors consisting of relevant financial market observations to the sequence of positions in each asset for all time-steps. The trainable parameters $\theta$ are optimized to minimize the expected value of a loss function applied to the terminal hedging error
	obtained as a result of the trading decisions made by the LSTM. The following definition describes more formally how the LSTM computes the hedging strategy. Note that in the numerical experiments of \cref{sec:numerical_results}, the hedging instruments used for the risk minimization of $\Phi$ are either only the underlying or standard options. The case of using both the underlying and options is not considered because of its redundancy; the options can replicate positions in the underlying with calls and puts.  
	
%
%
	%
	
	\begin{Def}{(Hedging with an LSTM)}
	\label{def:LSTM_hedging}
	Let $F_{\theta}$ be an LSTM as in \cref{def:LSTM} which maps the sequence of feature vectors $\{X_{t_{n}}\}_{n=0}^{N-1}$ to the output vectors $\{Y_{t_{n}}\}_{n=0}^{N-1}$. The choice of hedging instruments (i.e. the underlying or standard options) implies differences for the feature vectors and output vectors\footnote{
		%
		The computation of $\{V_{t_{n}}^{\delta}\}_{n=0}^{N-1}$ can be done for instance as in \eqref{eq_ref:portfolio_value_func} where asset positions are given by the output vectors of the LSTM.
	}:
	\begin{itemize}
		\item [1)] Hedging only with the underlying: the feature vector at each time-step is\footnote{
			Using the transformations 
			$\{\log(S_{t_{n}}^{(0,b)}), \log(Z_{t_{n}}), V_{t_{n}}^{\delta}/V_0^{\delta}\}$
			instead of $\{S_{t_{n}}^{(0,b)}, Z_{t_{n}}, V_{t_{n}}^{\delta}\}$ in feature vectors for the numerical experiments of \cref{sec:numerical_results} was found to significantly improve the training of neural networks. We note that the $\log$ transformation could not be applied for the hedging portfolio values since $V_{t_{n}}^{\delta}$ can theoretically take values on the real line.
		} 
		$$X_{t_{n}}:=[\log(S_{t_{n}}^{(0,b)}), \log(Z_{t_{n}}), V_{t_{n}}^{\delta}/V_0^{\delta}], \quad n = 0,\ldots,N-1,$$
		and $F_{\theta}$ outputs at each rebalancing date the position in the underlying: $\delta_{t_{n}}^{(0)} = Y_{t_{n-1}}$. 
		\item [2)] Hedging only with options: the feature vector at each time-step includes option prices as well as the price of the underlying: 
		$$X_{t_{n}}:=[\log(\bar{S}_{t_{n}}^{(b)}), \log(Z_{t_{n}}), V_{t_{n}}^{\delta}/V_0^{\delta}], \quad n = 0,\ldots,N-1,$$ 
		and $F_{\theta}$ outputs at each rebalancing date the position in the $D$ options: $[\delta_{t_{n}}^{(1)}, \ldots, \delta_{t_{n}}^{(D)}] = Y_{t_{n-1}}$.
	\end{itemize}	
	\end{Def}
	It is important to note that the choice of dynamics for the financial market could imply that relevant necessary information to compute the time-$t_{n}$ trading strategy should be added to feature vectors. For instance, \cite{carbonneau2020} apply the deep hedging algorithm with GARCH models which entails adding the volatility process to feature vectors. In the current paper, the models considered for the underlying imply that $\{S_{t_{n}}^{(0,b)}\}_{n=0}^{N}$ is a Markov process under $\mathbb{P}$ and thus that no additional variables must be added to feature vectors. Nevertheless, we note that the same methodological approach for hedging described in this section can easily be adapted to dynamics requiring the inclusion of additional state variables.
	
	\begin{remark}
		\label{remark:keep_Z_process}
		\cite{buehler2019deep_2} deep hedge exotic derivatives with an LSTM with feature vectors that does not 
		include a path-dependent state variable such as $\{Z_{t_{n}}\}_{n=0}^{N-1}$.
		The author of the current paper observed that adding $\{Z_{t_{n}}\}_{n=0}^{N-1}$ to feature vectors as per \cref{def:LSTM_hedging} significantly improved the performance of the optimized hedging policies when the number of trading period was large (i.e. for large $N$), while for less frequent trading, the gain was marginal.
	\end{remark}

	\begin{remark}
		Theoretical results from \cite{buehler2019deep} show that a FFNN
		could have been used to approximate arbitrarily well the optimal hedging policy in the setup considered in this study (see Proposition 4.3 of their paper). However, the author of the current paper observed that hedging with an LSTM was significantly more effective than with a FFNN for the numerical experiments conducted in \cref{sec:numerical_results} in terms of both computational time (i.e. faster learning with LSTMs) and hedging effectiveness which motivated the use of LSTMs as trading policies. The justifications of the superiority of LSTMs over FFNNs in the context of this paper are out-of-scope and are left out as interesting potential future work.
	\end{remark}

	For the rest of the paper, a single set of hyperparameters for the LSTM is considered in terms of the number of LSTM cells and neurons per cell.\footnote{
	Note that as per \cref{def:LSTM_hedging}, the dimensions of the input and output of the LSTM at each time-step, i.e. $d_{\text{in}}$ and $d_{\text{out}}$, are dependent of the choice of hedging instruments. Thus, while the number of neurons $d_{1},\ldots,d_{H}$ and the number of LSTM cells $H$ is fixed for the numerical experiments of \cref{sec:numerical_results}, the total number of trainable parameters will vary with respect to the choice of hedging instruments. 
	} The optimization problem thus consists in searching for the optimal values of trainable parameters for this specific architecture of LSTM. The hyperparameter tuning step is not considered in this paper; the reader is referred to \cite{buehler2019deep} or \cite{carbonneau2020} for a complete description of the optimal hedging problem with FFNNs which includes hyperparameter tuning.
	
	\begin{Def}{(Global risk exposure with an LSTM)}
	 Define $\tilde{\epsilon}(V_0)$ as the global risk exposure of the short position in $\Phi$ under optimal hedge if the hedging strategy is given by $F_{\theta}$ and if the value of the initial hedging portfolio is $V_{0} \in \mathbb{R}$:
	\begin{align}
	\tilde{\epsilon}(V_0) &:= \underset{\theta \in \mathbb{R}^{q}}{\min} \, \E\left[\mathcal{L}\left(\Phi(S_{T}^{(0,b)},Z_{T}) - V_{T}^{\delta^{\theta}}\right)  \right], \label{eq:risk_short_LSTM}
	\end{align}
	where $\delta^{\theta}$ is to be understood as the output vectors of $F_{\theta}$ and $q \in \mathbb{N}$ is the total number of trainable parameters.
	\end{Def}	

	\subsection{Training of neural networks}
	\label{subsec:training}
	The numerical scheme to optimize the trainable parameters $\theta$ is now described. For convenience, a similar notation as in the work of \cite{carbonneau2020} is used. For a given loss function and an initial portfolio value, the objective is to find $\theta$ such that the risk exposure of a short position in $\Phi$ is minimized (i.e. as in \eqref{eq:risk_short_LSTM}). The training procedure was originally proposed in \cite{buehler2019deep} and relies on (mini-batch) stochastic gradient descent (SGD), a very popular algorithm in the deep learning literature to train neural networks. Denote $J(\theta)$ as the cost function to minimize:
	$$J(\theta) := \E\left[\mathcal{L}\left(\Phi(S_{T}^{(0,b)},Z_{T}) - V_{T}^{\delta^{\theta}}\right)  \right], \quad \theta \in \mathbb{R}^{q}.$$
	Let $\theta_{0}$ be the initial values for the trainable parameters.\footnote{
	In this paper, the initial values of $\theta$ are always set as the \textit{glorot initialization} of \cite{glorot2010understanding}. 
	} The optimization procedure consists in the following iterations:
	\begin{align}
	\theta_{j+1} = \theta_{j} - \eta_{j} \nabla_{\theta} J(\theta_{j}), \label{eq:ref_SGD_step}
	\end{align}
	where $\nabla_{\theta}$ is the gradient operator with respect to $\theta$ and $\{\eta_{j}\}_{j\geq 0}$ is a sequence of small positive real values. In the context of this paper, $\nabla_{\theta} J(\theta)$ is unknown analytically and is estimated with Monte Carlo sampling. 
	Let $\mathbb{B}_{j}:=\{\pi_{i,j}\}_{i=1}^{N_{\text{batch}}}$ be a mini-batch of simulated hedging errors of size $N_{\text{batch}} \in \mathbb{N}$ with $\pi_{i,j}$ as the $i^{\text{th}}$ hedging error if $\theta = \theta_{j}$:
	$$\pi_{i,j}:=\Phi(S_{T,i}^{(0,b)},Z_{T,i}) - V_{T,i}^{\delta^{\theta_{j}}},$$
	where $S_{T,i}^{(0,b)},Z_{T,i}$ and $V_{T,i}^{\delta^{\theta_{j}}}$ are to be understood as the values of the $i^{\text{th}}$ simulated path. Moreover, denote
	%
	$\hat{J} : \mathbb{R}^{N_{\text{batch}}} \rightarrow \mathbb{R}$ as the empirical estimator of $J(\theta_{j})$ evaluated with $\mathbb{B}_{j}$ and $\nabla_{\theta} \hat{J}(\mathbb{B}_{j})$ as the empirical estimator of $\nabla_{\theta} J(\theta_{j})$ evaluated at $\theta = \theta_{j}$. In \cref{sec:numerical_results}, the MSE and SMSE penalties defined respectively as $\mathcal{L}^{\text{MSE}}(x):=x^{2}$ and $\mathcal{L}^{\text{SMSE}}(x):=x^{2} \mathds{1}_{\{x > 0\}}$ are extensively used. The empirical estimator of the cost function under each penalty can be stated as follows:
	\begin{align}
	\hat{J}^{\text{MSE}}(\mathbb{B}_{j})&:= \frac{1}{N_{\text{batch}}}\sum_{i=1}^{N_{\text{batch}}} \pi_{i,j}^{2}, \nonumber
	\\ \hat{J}^{\text{SMSE}}(\mathbb{B}_{j})&:= \frac{1}{N_{\text{batch}}}\sum_{i=1}^{N_{\text{batch}}} \pi_{i,j}^{2} \mathds{1}_{\{\pi_{i,j} > 0\}}. \label{eq:ref_SMSE_statistic}
	\end{align}
	One essential property of the architecture of neural networks is that the gradient of empirical cost functions (i.e. $\nabla_{\theta} \hat{J}(\mathbb{B}_{j})$ for both penalties) is known analytically. Indeed, we note that hedging errors are linearly dependent of the trading strategies produced as the outputs of the LSTM. Furthermore, the gradient of the outputs of an LSTM with respect to trainable parameters is known analytically (see e.g. Chapter 10 of \cite{goodfellow2016deep}).
	
	\begin{remark}
		In practice, the algorithm backpropagation through time (BPTT) is often used to compute analytically the gradient of a cost function with respect to the trainable parameters for recurrent type of neural networks such as an LSTM. BPTT leverages the structure of LSTMs (e.g. parameters sharing at each time-step) as well as the chain rule of calculus to obtain such gradients. In practice, efficient deep learning libraries such as Tensorflow \citep{abadi2016tensorflow} are often used to implement BPTT. Moreover, algorithms such as Adam \citep{kingma2014adam} which dynamically adapt the terms $\{\eta_{j}\}_{j \geq 0}$ in \eqref{eq:ref_SGD_step} have been shown to improve the training of neural networks. For the rest of the paper, Tensorflow and Adam are used to train every neural network.
	\end{remark}
	
	
	\section{Numerical study}
	\label{sec:numerical_results}
	In this section, an extensive numerical study benchmarking different dynamic hedging strategies for the long-term lookback option is presented. \cref{subsec:benchmark_hedging} benchmarks two global hedging strategies optimized with the deep hedging algorithm and the local risk minimization scheme of \cite{coleman2007robustly} with different hedging instruments and different dynamics for the financial market. \cref{subsec:policies_study} provides insight into specific characteristics of the optimized global policies. 
	The setup for the latter numerical experiments is described in \cref{subsec:numerical_procedure} and \cref{subsec:dynamics_market}.

	\subsection{Market setup}
	\label{subsec:numerical_procedure}
	The market setup considered in this paper is very similar to the work of \cite{coleman2007robustly}. The contingent claim to hedge is a lookback option of payoff $\Phi$ as in \eqref{eq:ref_payoff_lookback_option} with a time-to-maturity of $10$ years (i.e. $T=10$). The annualized continuous risk-free rate is set at $3\%$ (i.e. $r=0.03$) and $S_{0}^{(0,b)}=100$. In the design of hedging policies, the trading instruments considered are either the underlying, two options or six options. All options have a time-to-maturity of $1$ year, are traded once and are held until expiration. For the case of two options, the hedging instruments available at the beginning of each year consist of at-the-money (ATM) calls and puts. With six options, three calls of moneynesses $K \in \{S_{t_{n}}, 1.1S_{t_{n}}, 1.2S_{t_{n}}\}$ and three puts of moneynesses $K \in \{S_{t_{n}}, 0.9S_{t_{n}}, 0.8S_{t_{n}}\}$ are available at the beginning of each year $t_{n}$.
	As for the underlying, both monthly and yearly rebalancing are considered in numerical experiments. Yearly time-steps are used for all hedging instruments (i.e. $N=10$) except when hedging is done with the underlying on a monthly basis (i.e. $N=120$). 
	\begin{remark} 
		The methodological approach of \cref{sec:Methodology} is in no way dependent on this choice of hedging instruments.
	\end{remark}
	
	
	\subsubsection{Global hedging penalties}
	\label{subsubsec:global_hedging_penalties}
	The penalties studied for global hedging are the MSE and SMSE, and the respective optimization procedures are referred to as quadratic deep hedging (QDH) and semi-quadratic deep hedging (SQDH). While the MSE penalizes equally hedging gains and losses, the SMSE is more in line with the actual objectives of the hedger as it corresponds to an agent who strictly penalizes hedging losses proportionally to their squared values. 
	It is important to note that the computational cost of the deep hedging algorithm is closed to invariant to the choice of loss function. The motivation for assessing the effectiveness of QDH is the popularity of the quadratic penalty in the global hedging literature. 


	\subsubsection{LSTM training}
	\label{subsubsec:hyperparameters} 
	The training of the LSTM is done as described in \cref{subsec:training} on a training set of $350,\!000$ paths with $150$ epochs\footnote{
		One epoch is defined as a complete iteration of SGD on the training set. For a training set and mini-batch size of respectively $350,\!000$ and $1,\!000$, one epoch consists of a total of $350$ updates of parameters as in \eqref{eq:ref_SGD_step}.
	} and a mini-batch size of $1,\!000$. A validation set of $75,\!000$ paths is used to find the optimal set of trainable parameters out of the $150$ epochs. More precisely, 
	at the end of each epoch, the hedging metric associated to the penalty being optimized (i.e. MSE for QDH and SMSE for SQDH) is evaluated on the validation set at the current values of the trainable parameters. The optimal set of trainable parameters is approximated by the one that minimizes the empirical cost function on the validation set out of $150$ epochs. The use of a validation set to select the number of epochs was found to significantly improve the out-of-sample hedging performance obtained with SQDH, while for QDH, the improvement was marginal. 
	
	All results presented in subsequent sections are from a test set (out-of-sample) of $75,\!000$ paths. The structure of the LSTM is as in \cref{def:LSTM} with two LSTM cells (i.e. $H = 2$) and $24$ neurons per cell (i.e. $d_1=d_2=24$). The Adam optimizer (\cite{kingma2014adam}) is used for all examples with a learning rate of $0.01$ for QDH and $\frac{0.01}{6}$ for SQDH since a smaller learning rate was found to improve the training under the SMSE penalty.
	
	\subsubsection{Local risk minimization}
	 Define $\{C_{t_{n}}^{\delta}\}_{n=0}^{N}$ as the discounted cumulative cost process associated to a trading strategy $\delta$: 
	$$C_{t_{n}}^{\delta}:=B_{t_{n}}^{-1}V_{t_{n}}^{\delta} - G_{t_{n}}^{\delta}, \quad n = 0,\ldots,N.$$
	Contrarily to global hedging, local risk minimization results in strategies that are not necessarily self-financing. Indeed, the optimization of hedging strategies under this framework imposes the constraint that the terminal portfolio value exactly matches the payoff of the contingent claim, i.e. $V_{T}^{\delta} = \Phi(S_{T}^{(0,b)},Z_{T})$ $\mathbb{P}$-a.s., which can always be respected by the injection or withdrawal of capital at time $T$. Under this constraint, local risk minimization optimizes at each time-step starting backward from time $T$ positions in the assets which minimize the expected squared incremental cost. More precisely, for $n = N-1,\ldots,0$, the optimization aims at finding $(\delta_{t_{n+1}}^{(0:D)}, \delta_{t_{n+1}}^{(B)})$ that minimize $\E[(C_{t_{n+1}}^{\delta} - C_{t_{n}}^{\delta})^{2}|\mathcal{F}_{t_{n}}]$ at time $t_{n}$ with the constraint that $V_{T}^{\delta} = \Phi(S_{T}^{(0,b)},Z_{T})$ $\mathbb{P}$-a.s. The optimal initial capital amount to invest denoted as $V_{0}^{\star}$ is also obtained as a result of this scheme. Once the trading strategy $\delta$ is optimized with the local risk minimization procedure, a self-financing strategy can be constructed by setting the initial portfolio value as $V_{0}^{\delta}=V_{0}^{\star}$, by following the optimized trading strategy strictly for the risky assets (i.e. $\delta_{t_{n}}^{(0:D)}$ for $n=1,\ldots,N$) and by adjusting positions in the risk-free asset such that the trading strategy is self-financing (i.e. respecting \eqref{eq:ref_self_financing}). Hedging results presented in the numerical experiments of this section with local risk minimization are self-financing as per the latter description and are from the work of \cite{coleman2007robustly}. For examples of numerical schemes to implement local risk procedures, the reader is referred to \cite{coleman2006hedging} or \cite{augustyniak2017assessing}.  
	
	The motivation for benchmarking the global policies optimized with our methodological approach to local risk minimization is twofold. First, local risk procedures are popular for the risk mitigation of VAs guarantees in the literature (e.g. \cite{coleman2006hedging}, \cite{coleman2007robustly}, \cite{kelani2017pricing}, \cite{trottier2018fund} and \cite{trottier2018local}). 
	Second, in the context of hedging European vanilla options of maturity one to three years, \cite{augustyniak2017assessing} showed that global quadratic hedging with the underlying improves upon the downside risk reduction over local risk minimization. The question remains if the latter holds for longer maturities and when liquid options are used as hedging instruments.
	
%
	\subsubsection{Hedging metrics}
	The hedging metrics considered for the benchmarking of the different trading policies include the root-mean-square error (RMSE) and the semi-RMSE (i.e. the root of the SMSE statistic). Tail risk metrics are also studied with the Value-at-Risk (VaR) and the Conditional Value-at-Risk (CVaR, \cite{rockafellar2002conditional}). For an absolutely continuous integrable random variable\footnote{
		All dynamics assumed for the underlying in \cref{sec:numerical_results} imply that hedging errors are absolutely continuous integrable random variables.
	}, the CVaR at confidence level $\alpha$ has the following representation:
	\begin{align}
	\text{CVaR}_{\alpha}(X) := \E[X|X \geq \text{VaR}_{\alpha}(X)], \quad \alpha \in (0,1), \label{eq:ref_CVaR}
	\end{align}
	where $\text{VaR}_{\alpha}(X):= \underset{x}{\min} \, \{x|\mathbb{P}(X \leq x) \geq \alpha\}$ is the VaR at confidence level $\alpha$. The $\text{CVaR}_{\alpha}$ represents tail risk by averaging all hedging errors larger than the $\alpha^{\text{th}}$ percentile of the distribution of hedging errors (i.e. the $\text{VaR}_{\alpha}$ metric). Hedging statistics presented in subsequent sections are estimated with conventional empirical estimators on the test set.

	
	\subsection{Dynamics of financial market}
	\label{subsec:dynamics_market}
	The choice of dynamics for the underlying is motivated by the objective of studying the optimized global policies under different stylized features of the financial market. It is important to recall that deep hedging is a model-free reinforcement learning approach: the LSTM is never explicitly told the dynamics of the financial market during its training phase. Instead, the neural network must learn through many simulations of a market generator how to dynamically adapt its embedded policy, i.e. its trainable parameters, with the objective of minimizing the expected loss function of the resulting hedging errors. The current work studies the impact of the presence of jump risk on optimized global policies by considering the Merton jump-diffusion model (MJD, \cite{Merton1976}) as well as the  Black-Scholes model (BSM, \cite{black1973pricing}). 
	Both dynamics are described subsequently and the parameters values presented in \cref{table:params_BSM} and \cref{table:params_MJD}
	are the same as in \cite{coleman2007robustly}. It is worth noting that while the values of the parameters imply somewhat similar periodic means and standard deviations for log-returns, the MJD parameters entail large and volatile negative jumps occurring on average once over the lifetime of the lookback option.
	
	Moreover, the stochastic models considered in this paper imply that the market is arbitrage-free. By the first fundamental theorem of asset pricing, there exist a probability measure $\mathbb{Q}$ equivalent to $\mathbb{P}$ such that $\{e^{-r t_n}S_{t_{n}}^{(b,0)}\}_{n=0}^{N}$ is an ($\mathbb{F}, \mathbb{Q}$)-martingale (see, for instance, \cite{delbaen1994general}). Let $y_{t_{n}}:=\log(S_{t_{n}}^{(0,b)}/S_{t_{n-1}}^{(0,b)})$ be the periodic log-return of the underlying, and $\{\epsilon_{t_{n}}^{\mathbb{P}}\}_{n=1}^{N}$ and $\{\epsilon_{t_{n}}^{\mathbb{Q}}\}_{n=1}^{N}$ be sequences of independent standard normal random variables under respectively $\mathbb{P}$ and $\mathbb{Q}$. The dynamics of both models are now formally defined.
	
	\subsubsection{BSM under $\mathbb{P}$}
	\label{subsec:BSM_model_under_P}
	The discrete BSM assumes that log-returns are i.i.d. $\!\!$ normal random variables of periodic mean and variance of respectively $(\mu-\frac{\sigma^{2}}{2})\Delta_{N}$ and $\sigma^{2} \Delta_{N}$:
	\begin{align}
	y_{t_{n}}&=\left(\mu-\frac{\sigma^{2}}{2}\right)\Delta_{N} + \sigma \sqrt{\Delta_{N}}\epsilon_{t_{n}}^{\mathbb{P}}, \quad n = 1,\ldots, N, \label{eq:ref_BSM_under_P}
	\end{align}
	where $\mu \in \mathbb{R}$ and $\sigma > 0$ are the yearly model parameters. 
	
	
	\subsubsection{MJD under $\mathbb{P}$}
	\label{subsec:MJD_model_under_P}
	The MJD model extends the BSM by assuming the presence of random jumps to the underlying stock price. More precisely, let $\{\zeta_k^{\mathbb{P}}\}_{k=1}^{\infty}$ be independent normal random variables of mean $\mu_J$ and variance $\sigma_J^{2}$, and $\{N_{t_{n}}^{\mathbb{P}}\}_{n=0}^{N}$ be values of a Poisson process of intensity $\lambda > 0$ where $\{\zeta_k^{\mathbb{P}}\}_{k=1}^{\infty}, \{N_{t_{n}}^{\mathbb{P}}\}_{n=0}^{N}$ and $\{\epsilon_{t_{n}}^{\mathbb{P}}\}_{n=1}^{N}$ are independent. Periodic log-returns under this model can be stated as follows\footnote{
		We adopt the convention that if $N_{t_{n}}^{\mathbb{P}}=N_{t_{n-1}}^{\mathbb{P}}$, then:
		$$\sum_{k=N_{t_{n-1}}^{\mathbb{P}}+1}^{N_{t_{n}}^{\mathbb{P}}}\zeta_{k}^{\mathbb{P}} = 0.$$ 
	}: 
	\begin{align}
	y_{t_{n}} = \left(\alpha - \lambda \left(e^{\mu_J + \sigma_J^{2}/2}-1\right) -  \frac{\sigma^{2}}{2}\right)\Delta_{N} + \sigma \sqrt{\Delta_{N}}\epsilon_{t_{n}}^{\mathbb{P}} + \sum_{k=N_{t_{n-1}}^{\mathbb{P}} + 1}^{N_{t_{n}}^{\mathbb{P}}}\zeta_{k}^{\mathbb{P}}, \label{eq:ref_MJD_under_P}
	\end{align}
	where $\{\alpha, \mu_J, \sigma_J, \lambda, \sigma\}$ are the model parameters with $\{\alpha, \lambda, \sigma\}$ being on a yearly scale, $\alpha \in \mathbb{R}$ and $\sigma > 0$. 
	
	
	\subsubsection{BSM under $\mathbb{Q}$}
	\label{subsec:BSM_model_under_Q}
	By a discrete-time version of the Girsanov theorem, there exist an $\mathbb{F}$-adapted market price of risk process $\{\varphi_{t_{n}}\}_{n=1}^{N}$ such that 
	\begin{align}
	\epsilon_{t_{n}}^{\mathbb{Q}} = \epsilon_{t_{n}}^{\mathbb{P}} - \varphi_{t_{n}}, \quad n = 1,\ldots,N. \label{eq:ref_market_price_risk_BSM}
	\end{align}
	For $n=1,\ldots,N$, let $\varphi_{t_{n}} := -\sqrt{\Delta_{N}}\left(\frac{\mu-r}{\sigma}\right)$. By replacing $\epsilon_{t_{n}}^{\mathbb{P}} =  \epsilon_{t_{n}}^{\mathbb{Q}} + \varphi_{t_{n}}$ into \eqref{eq:ref_BSM_under_P}, it is straightforward to obtain the $\mathbb{Q}$-dynamics of log-returns:
	\begin{align}
	y_{t_{n}} &= \left(r-\frac{\sigma^{2}}{2}\right)\Delta_{N} + \sigma \sqrt{\Delta_{N}} \epsilon_{t_{n}}^{\mathbb{Q}}, \quad n = 	1,\ldots, N. \label{eq:ref_BSM_under_Q}
	\end{align}
	The pricing of European calls and puts used as hedging instruments under this model is done with the well-known Black-Scholes closed-form solutions.
	
	
	\subsubsection{MJD under $\mathbb{Q}$}
	\label{subsec:MJD_model_under_Q}
	The change of measure considered is the same as the one from \cite{coleman2007robustly}. Let $\{\zeta_{k}^{\mathbb{Q}}\}_{k=1}^{\infty}$ be independent normal random variables under $\mathbb{Q}$ of mean $\tilde{\mu}_J$ and variance $\tilde{\sigma}_J^{2}$, and $\{N_{t_{n}}^{\mathbb{Q}}\}_{n=0}^{N}$ be values of a Poisson process of intensity $\tilde{\lambda}>0$ where $\{\zeta_{k}^{\mathbb{Q}}\}_{k=1}^{\infty}$, $\{N_{t_{n}}^{\mathbb{Q}}\}_{n=0}^{N}$ and $\{\epsilon_{t_{n}}^{\mathbb{Q}}\}_{n=1}^{N}$ are independent. The $\mathbb{Q}$-dynamics of log-returns can be stated as follows:
	\begin{align}
	y_{t_{n}} = \left(r - \tilde{\lambda} \left(e^{\tilde{\mu}_J + \tilde{\sigma}_J^{2}/2}-1\right) -  \frac{\sigma^{2}}{2}\right)\Delta_{N} + \sigma \sqrt{\Delta_{N}}\epsilon_{t_{n}}^{\mathbb{Q}} + \sum_{k=N_{t_{n-1}}^{\mathbb{Q}} + 1}^{N_{t_{n}}^{\mathbb{Q}}}\zeta_{k}^{\mathbb{Q}}, \nonumber
	\end{align} 
	where $\tilde{\sigma}_J:=\sigma_J$, $\tilde{\mu}_J:= \mu_J - (1-\gamma)\sigma_J^{2}, \tilde{\lambda}:=\lambda e^{-(1-\gamma)(\mu_J-\frac{1}{2}(1-\gamma)\sigma_J^{2})}$ with $\gamma \leq 1$ as the risk aversion parameter which is set at $\gamma = -1.5$. The value of the risk aversion parameter implies more frequent and more negative jumps on average under $\mathbb{Q}$ than under $\mathbb{P}$ by increasing $\tilde{\lambda}$ and decreasing $\tilde{\mu}_J$.
	The pricing of European calls and puts used as hedging instruments under the MJD model is done with the well-known closed-form solutions.

	
	\begin{table}[!htbp]
		\caption {Parameters of the Black-Scholes model.} \label{table:params_BSM}
		\begin{adjustwidth}{-1in}{-1in} 
			\centering
			\begin{tabular}{cc}
				\hline
				$\mu$ & $\sigma$
				\\
				\hline\noalign{\medskip}
				$0.10$  &  $0.15$
				\\    
				\noalign{\medskip}\hline
			\end{tabular}%
		\end{adjustwidth}
		\centering{Notes: Both $\mu$ and $\sigma$ are on an annual basis.}
	\end{table}

	\begin{table}[!htbp]		
		\caption {Parameters of the Merton jump-diffusion model.}  \label{table:params_MJD}
		\begin{adjustwidth}{-1in}{-1in} 
			\centering  
			\begin{tabular}{cccccc}
				\hline
				$\alpha$ & $\sigma$ & $\lambda$ & $\mu_{J}$ & $\sigma_{J}$ & $\gamma$ 
				\\
				\hline\noalign{\medskip}
				$0.10$  &  $0.15$ & $0.10$ & $-0.20$ & $0.15$ & $-1.5$
				\\    
				\noalign{\medskip}\hline
			\end{tabular}%
		\end{adjustwidth}
		\centering{Notes: $\alpha$, $\sigma$ and $\lambda$ are on an annual basis.}
	\end{table}
	

	\subsection{Benchmarking of hedging policies}
	\label{subsec:benchmark_hedging}
	In this section, the hedging effectiveness of QDH, SQDH and local risk minimization is assessed under various market settings. The analysis starts off in \cref{subsubsec:results_local_global} by comparing QDH and local risk minimization performance as both approaches are optimized with a quadratic criterion; the benchmarking of global hedging policies embedded in QDH and SQDH is done in \cref{subsubsec:QDH_vs_SQDH}. 

	
	\subsubsection{QDH and local risk minimization benchmark} 
	\label{subsubsec:results_local_global}
	\cref{table:BSM_Coleman_2007_MSE_vs_local} and \cref{table:MJD_Coleman_2007_MSE_vs_local} presents hedging statistics of QDH and local risk minimization under respectively the BSM and MJD model.\footnote{
	The choice of hedging statistics presented in \cref{table:BSM_Coleman_2007_MSE_vs_local} and \cref{table:MJD_Coleman_2007_MSE_vs_local} are the ones considered in \cite{coleman2007robustly}. Additional hedging statistics for QDH are presented in \cref{subsubsec:QDH_vs_SQDH}.
	} For comparative purposes, the initial capital investment 
	is set to the optimized value obtained as a result of the local risk minimization procedure of \cite{coleman2007robustly} for all examples. We note that this choice naturally gives a disadvantage to QDH.
	\begin{table}[ht]
	\caption {Benchmarking of quadratic deep hedging (QDH) and local risk minimization to hedge the lookback option of $T = 10$ years under the BSM.} 
	\label{table:BSM_Coleman_2007_MSE_vs_local}
	\renewcommand{\arraystretch}{1.15}
	\begin{adjustwidth}{-1in}{-1in} 
		\centering
		\begin{tabular}{lccccccccc}
			\hline\noalign{\smallskip}
			& & \multicolumn{3}{c}{$\text{Local risk minimization}$} & & \multicolumn{3}{c}{$\text{QDH}$} & \\
			\cline{3-5}\cline{7-9}   Statistics  & $V_{0}^{\delta}$ & RMSE & $\text{VaR}_{0.95}$ & $\text{CVaR}_{0.95}$ &  & RMSE & $\text{VaR}_{0.95}$ & $\text{CVaR}_{0.95}$ &  \\
			\hline\noalign{\medskip} 
			%
			Stock (year)  & $13.9$ & $15.9$ & $28.5$ & $43.2$ &   & $14.8$ & $25.5$   & $41.2$ &    \\
			Stock (month) & $17.3$ & $5.5$  & $8.9$  & $13.0$ &   & $4.9$  & $7.7$   & $12.2$ &    \\
			Two options   & $17.4$ & $4.6$  & $7.0$  & $11.9$ &   & $4.2$  & $6.1$   & $11.2$ &    \\
			Six options   & $17.7$ & $1.6$  & $2.4$  & $3.8$  &   & $1.1$  & $1.2$   & $2.4$ &    \\
			%
			\noalign{\medskip}\hline
		\end{tabular}%
	\end{adjustwidth}
	Notes: Hedging statistics under the BSM with $\mu = 0.1, \sigma = 0.15, r = 0.03$ and $S_0^{(0,b)} = 100$ (see \cref{subsec:BSM_model_under_P} for model description under $\mathbb{P}$ and \cref{subsec:BSM_model_under_Q} for the risk-neutral dynamics used for option pricing). \textit{Hedging instruments}: monthly and yearly underlying, yearly ATM call and put options (\textit{two options}) and three yearly calls and puts of strikes $K = \{S_{t_{n}}^{(0,b)}, 1.1S_{t_{n}}^{(0,b)}, 1.2S_{t_{n}}^{(0,b)}\}$ and $K = \{S_{t_{n}}^{(0,b)}, 0.9S_{t_{n}}^{(0,b)}, 0.8S_{t_{n}}^{(0,b)}\}$ (\textit{six options}). Results for local risk minimization and initial portfolio values $V_{0}^{\delta}$ are from Table $3$ of \cite{coleman2007robustly}. Results for QDH are computed based on $75,\!000$ independent paths generated from the BSM under $\mathbb{P}$. Training of the neural networks is done as described in \cref{subsubsec:hyperparameters}. 
	\end{table}
	\begin{table}[ht]
		\caption {Benchmarking of quadratic deep hedging (QDH) and local risk minimization to hedge the lookback option of $T = 10$ years under the MJD model.}  
		\label{table:MJD_Coleman_2007_MSE_vs_local}
		\renewcommand{\arraystretch}{1.15}
		\begin{adjustwidth}{-1in}{-1in} 
			\centering
			\begin{tabular}{lccccccccc}
				\hline\noalign{\smallskip}
				& & \multicolumn{3}{c}{$\text{Local risk minimization}$} & & \multicolumn{3}{c}{$\text{QDH}$} & \\
				\cline{3-5}\cline{7-9}   Statistics  & $V_{0}^{\delta}$ & RMSE & $\text{VaR}_{0.95}$ & $\text{CVaR}_{0.95}$ &  & RMSE & $\text{VaR}_{0.95}$ & $\text{CVaR}_{0.95}$ &  \\
				\hline\noalign{\medskip} 
				%
				Stock (year)  & $19.5$ & $21.4$  & $38.4$  & $60.5$ &   & $19.5$ & $33.1$ & $55.8$   &    \\
				Stock (month) & $22.8$ & $13.0$  & $23.5$  & $38.4$ &   & $11.0$ & $16.3$ & $33.5$   &    \\
				Two options   & $24.6$ & $6.0$   & $8.4$   & $15.2$ &   & $5.2$ & $6.7$ & $12.9$   &    \\
				Six options   & $25.2$ & $1.9$   & $2.8$   & $4.6$  &   & $1.3$ & $1.7$ & $3.2$   &    \\
				\noalign{\medskip}\hline
			\end{tabular}%
		\end{adjustwidth}
		Notes: Hedging statistics under the MJD model with $\alpha = 0.1, \sigma = 0.15, \lambda =0.1, \mu_J = -0.2, \sigma_J = 0.15, \gamma = -1.5, r = 0.03$ and $S_{0}^{(0,b)} = 100$ (see \cref{subsec:MJD_model_under_P} for model description under $\mathbb{P}$ and \cref{subsec:MJD_model_under_Q} for the risk-neutral dynamics used for option pricing). \textit{Hedging instruments}: monthly and yearly underlying, yearly ATM call and put options (\textit{two options}) and three yearly calls and puts of strikes $K = \{S_{t_{n}}^{(0,b)}, 1.1S_{t_{n}}^{(0,b)}, 1.2S_{t_{n}}^{(0,b)}\}$ and $K = \{S_{t_{n}}^{(0,b)}, 0.9S_{t_{n}}^{(0,b)}, 0.8S_{t_{n}}^{(0,b)}\}$ (\textit{six options}). Results for local risk minimization and initial portfolio values $V_{0}^{\delta}$ are from Table $4$ of \cite{coleman2007robustly}. Results for QDH are computed based on $75,\!000$ independent paths generated from the MJD model under $\mathbb{P}$. Training of the neural networks is done as described in \cref{subsubsec:hyperparameters}. 
	\end{table}

	Since QDH optimizes the MSE penalty, the latter was expected to outperform local risk minimization on the RMSE metric. The question remained if QDH also improved upon the downside risk captured by the $\text{VaR}_{0.95}$ and $\text{CVaR}_{0.95}$ statistics. Numerical results under both dynamics demonstrate that QDH outperforms local risk minimization across all downside risk metrics and all hedging instruments. The risk reduction obtained with QDH over local risk minimization is most impressive with six options: the percentage decrease for respectively the RMSE, $\text{VaR}_{0.95}$ and $\text{CVaR}_{0.95}$ statistics are of $33\%, 52\%$ and $36\%$ under the BSM and of $27\%, 38\%$ and $30\%$ under the MJD model. As for hedging with the underlying on a monthly and yearly basis as well as with two options, the improvement of QDH over local risk minimization for the three hedging statistics 
	ranges between $5\%$ to $13\%$ under the BSM and $8\%$ to $20\%$ under the MJD model except for the $\text{VaR}_{0.95}$ metric with the stock on a monthly basis under the MJD dynamics which achieves $30\%$ reduction.
	These results demonstrate that the use of a global procedure rather than a local procedure provides better hedging performance.
%
	\subsubsection{QDH and SQDH benchmark}
	\label{subsubsec:QDH_vs_SQDH}
	The benchmarking of QDH and SQDH policies is now presented with the same setup as in the previous section except for the initial capital investment which is set as the risk-neutral price of the lookback option under both dynamics for all hedging instruments: $17.7\$$ for BSM and $25.3\$$ for MJD.\footnote{
	Risk-neutral prices of the lookback option were estimated with simulations for both dynamics.
	} This choice is motivated by the objective of comparing on common grounds the results obtained across the different hedging instruments for both global hedging approaches. \cref{table:BSM_MSE_vs_SMSE_no_TC} and \cref{table:MJD_MSE_vs_SMSE_no_TC} present descriptive statistics of the hedging shortfall obtained with QDH and SQDH under respectively the BSM and MJD model.
	\begin{table}[ht]
	\caption {Benchmarking of quadratic deep hedging (QDH) and semi-quadratic deep hedging (SQDH) to hedge the lookback option of $T = 10$ years under the BSM.} \label{table:BSM_MSE_vs_SMSE_no_TC}
	\renewcommand{\arraystretch}{1.15}
	\begin{adjustwidth}{-1in}{-1in} 
		\centering
		\begin{tabular}{lcccccccc}
		\hline\noalign{\smallskip}
		Statistics   & Mean & RMSE & semi-RMSE & $\text{VaR}_{0.95}$ & $\text{VaR}_{0.99}$ & $\text{CVaR}_{0.95}$ & $\text{CVaR}_{0.99}$ & Skew  \\
		\hline\noalign{\medskip} 
		\multicolumn{4}{l}{\underline{\emph{QDH}}} \\\noalign{\smallskip} 
		Stock (year)  & $-0.5$   & $14.8$  & $12.0$  & $25.8$  & $49.8$  & $41.4$  & $69.9$ & $1.9$ \\
		Stock (month) & $0.2$   & $4.9$  & $3.6$  & $7.9$  & $14.5$  & $12.1$  & $19.4$ & $0.4$ \\
		Two options   & $0.0$   & $4.2$  & $3.2$  & $6.5$  & $14.0$   & $11.5$   & $20.9$ & $1.9$ \\
		Six options   & $0.0$   & $1.1$  & $0.8$  & $1.2$  & $2.8$   & $2.4$   & $5.1$ & $3.3$ \\

		&   &   &  & & & &  \\
		\multicolumn{4}{l}{\underline{\emph{SQDH}}} \\\noalign{\smallskip} 
		%
		Stock (year)  & $-32.1$   & $43.8$ & $4.4$ & $6.4$ & $21.1$   & $16.0$   & $32.7$ & $-1.0$ \\
		Stock (month) & $-10.1$   & $15.0$  & $1.5$  & $2.5$  & $6.1$   & $4.9$   & $9.4$ & $-1.5$ \\
		Two options   & $-5.4$   & $10.0$  & $1.6$  & $1.3$  & $5.5$   & $4.1$   & $9.8$ & $-2.3$ \\
		Six options   & $-0.9$   & $2.2$  & $0.4$  & $0.2$  & $1.0$   & $0.8$   & $2.0$ & $-5.0$ \\
		\noalign{\medskip}\hline
	\end{tabular}%
	\end{adjustwidth}
	Notes: Hedging statistics under the BSM with $\mu = 0.1, \sigma = 0.15, r = 0.03, S_{0}^{(0,b)} = 100$ and $V_0^{\delta}=17.7$ for all examples (see \cref{subsec:BSM_model_under_P} for model description under $\mathbb{P}$ and \cref{subsec:BSM_model_under_Q} for the risk-neutral dynamics used for option pricing). \textit{Hedging instruments}: monthly and yearly underlying, yearly ATM call and put options (\textit{two options}) and three yearly calls and puts of strikes $K = \{S_{t_{n}}^{(0,b)}, 1.1S_{t_{n}}^{(0,b)}, 1.2S_{t_{n}}^{(0,b)}\}$ and $K = \{S_{t_{n}}^{(0,b)}, 0.9S_{t_{n}}^{(0,b)}, 0.8S_{t_{n}}^{(0,b)}\}$ (\textit{six options}). Results for each penalty are computed based on $75,\!000$ independent paths generated from the BSM under $\mathbb{P}$. Training of the neural networks is done as described in \cref{subsubsec:hyperparameters}. 	
	\end{table}
	\begin{table}[ht]
	\caption {Benchmarking of quadratic deep hedging (QDH) and semi-quadratic deep hedging (SQDH) to hedge the lookback option of $T = 10$ years under the MJD model.} \label{table:MJD_MSE_vs_SMSE_no_TC}
	\renewcommand{\arraystretch}{1.15}
	\begin{adjustwidth}{-1in}{-1in} 
	\centering
		\begin{tabular}{lcccccccc}
		\hline\noalign{\smallskip}
		Statistics   & Mean & RMSE & semi-RMSE & $\text{VaR}_{0.95}$ & $\text{VaR}_{0.99}$ & $\text{CVaR}_{0.95}$ & $\text{CVaR}_{0.99}$ & Skew \\
		\hline\noalign{\medskip} 
		\multicolumn{4}{l}{\underline{\emph{QDH}}} \\\noalign{\smallskip}  %
		Stock (year)  & $-1.6$   & $19.8$ & $15.6$ & $32.3$ & $66.4$   & $54.5$   & $95.4$ & $2.1$ \\
		Stock (month) & $0.2$   & $11.2$  & $9.4$  & $15.7$  & $42.8$   & $32.6$   & $64.6$ & $3.2$ \\
		Two options   & $0.0$   & $5.2$  & $3.8$  & $6.7$  & $15.4$   & $12.7$   & $25.1$ & $1.6$ \\
		Six options   & $-0.1$   & $1.3$  & $0.9$  & $1.4$  & $3.6$   & $2.9$   & $6.2$ & $2.3$ \\

		&   &   &  & & & &  \\
		\multicolumn{4}{l}{\underline{\emph{SQDH}}} \\\noalign{\smallskip} 
		%
		Stock (year)  & $-35.2$   & $49.7$ & $6.7$ & $11.4$ & $31.7$   & $24.6$   & $47.7$ & $-0.8$ \\
		Stock (month) & $-22.8$   & $33.8$  & $4.2$  & $6.5$  & $18.3$   & $14.3$   & $29.6$ & $-1.1$ \\
		Two options   & $-5.9$   & $11.2$  & $1.7$  & $2.2$  & $7.1$   & $5.5$   & $12.2$ & $-2.5$ \\
		Six options   & $-1.3$   & $3.1$  & $0.5$  & $0.3$  & $1.4$   & $1.1$   & $2.9$ & $-4.8$ \\
		\noalign{\medskip}\hline
		\end{tabular}%
	\end{adjustwidth}
	Notes: Hedging statistics under the MJD model with $\alpha = 0.1, \sigma = 0.15, \lambda =0.1, \mu_J = -0.2, \sigma_J = 0.15, \gamma = -1.5, r = 0.03, S_{0}^{(0,b)}=100$ and $V_0^{\delta} = 25.3$ for all examples (see \cref{subsec:MJD_model_under_P} for model description under $\mathbb{P}$ and \cref{subsec:MJD_model_under_Q} for the risk-neutral dynamics used for option pricing). \textit{Hedging instruments}: monthly and yearly underlying, yearly ATM call and put options (\textit{two options}) and three yearly calls and puts of strikes $K = \{S_{t_{n}}^{(0,b)}, 1.1S_{t_{n}}^{(0,b)}, 1.2S_{t_{n}}^{(0,b)}\}$ and $K = \{S_{t_{n}}^{(0,b)}, 0.9S_{t_{n}}^{(0,b)}, 0.8S_{t_{n}}^{(0,b)}\}$ (\textit{six options}). Results for each penalty are computed based on $75,\!000$ independent paths generated from the MJD model. Training of the neural networks is done as described in \cref{subsubsec:hyperparameters}.
	\end{table}
	Numerical results indicate that as compared to QDH, SQDH policies result in downside risk metrics two to three times smaller for almost all examples and earn significant gains across all hedging instruments (i.e. negative mean hedging errors). While QDH minimizes the RMSE statistic, the downside risk captured by the semi-RMSE, $\text{VaR}_{\alpha}$ and $\text{CVaR}_{\alpha}$ statistics for $\alpha$ equal to $0.95$ and $0.99$ are always significantly reduced by SQDH policies. Indeed, the downside risk reduction with SQDH over QDH in the latter hedging statistics ranges between $51\%$ to $85\%$ under the BSM and $45\%$ to $76\%$ under the MJD model.
	These impressive gains in risk reduction can be attributed to the fact that QDH penalizes equally upside and downside risk, 
	while on the other hand, SQDH strictly penalizes hedging losses proportionally to their squared values. 
	Furthermore, hedging statistics also indicate that SQDH policies achieve significant gains under both models and across all hedging instruments with a lesser extend for six options. We observe that hedging with the underlying on a yearly basis result in the most expected gains, followed by monthly underlying, two options and six options. All of these results clearly demonstrate that SQDH policies should be prioritized over QDH policies as they are tailor-made to match the financial objectives of the hedger by always significantly reducing the downside risk as well as earning positive returns on average. \cref{subsec:policies_study} that follows will shed some light on specific characteristics of the SQDH policies which result in these large average hedging gains and downside risk reduction.
	Moreover, it is also interesting to note that the distinct treatment of hedging shortfalls by each penalty has a direct implication on the skewness statistic. Indeed, by strictly optimizing squared hedging losses, SQDH effectively minimize the right tail of hedging errors which entails negative skewness. As for QDH, the positive skewness for all examples can be explained by the fact that the payoff of the lookback option is highly positively asymmetric since it is bounded below at zero and has no upper bound.

	Lastly, \cite{coleman2007robustly} observed with local risk minimization that while hedging with six options always results in better policies in terms of hedging effectiveness, the relative performance of using yearly ATM call and put options (i.e. two options) or the underlying on a monthly basis depends on the dynamics of the risky asset. The same conclusions can be made from our results obtained with global hedging. Indeed, hedging statistics of both QDH and SQDH policies under the Black-Scholes dynamics in \cref{table:BSM_MSE_vs_SMSE_no_TC} show that the downside risk metrics are most often only slightly better with two options as compared to hedging with the underlying on a monthly basis. On the other hand, values from \cref{table:MJD_MSE_vs_SMSE_no_TC} indicate that hedging with two options under the MJD model result in downside risk metrics at least two times smaller than with the underlying on a monthly basis for both QDH and SQDH. This observation stems from the fact that hedging with options is significantly more effective than with the underlying in the presence of jump risk. Thus, our results show that the observation made by \cite{coleman2007robustly} with respect to the significant improvement in hedging effectiveness of local risk minimization with options in the presence of jump risk also holds for both QDH and SQDH policies. 

    
    \subsection{Qualitative characteristics of global policies}
    \label{subsec:policies_study}
    While the previous section assessed the hedging performance of QDH and SQDH with various hedging instruments and different market scenarios, the current section provides insights into specific characteristics of the optimized global policies. The analysis starts off by comparing the \textit{average equity risk exposure} of QDH and SQDH policies, also  called \textit{average exposure} for convenience, with the same dynamics for the underlying as in previous sections (i.e. BSM and MJD model). The motivation of the latter is to assess if either the MSE or SMSE penalty result in hedging policies more geared towards being long equity risk and are thus earning the equity risk premium. 
    In this paper, the equity risk exposure is measured as the average portfolio delta over one complete path of the financial market. More formally, for $(\delta_{t_{n+1}}^{(0:D)}, \delta_{t_{n+1}}^{(B)})$ given and fixed, the portfolio delta at the beginning of year $t_{n}$ denoted as $\tilde{\Delta}^{(pf)}_{t_{n}}$ is defined as
	\begin{align}
	\tilde{\Delta}^{(pf)}_{t_{n}} &:= \frac{\partial V_{t_{n}}^{\delta}}{\partial S_{t_{n}}^{(0,b)}} \nonumber
	\\ &= \frac{\partial}{\partial S_{t_{n}}^{(0,b)}} \left(\delta_{t_{n+1}}^{(0:D)} \bigcdot \bar{S}_{t_{n}}^{(b)} + \delta_{t_{n+1}}^{(B)}B_{t_{n}}\right) \nonumber
	%
	%
	\\ &= \delta_{t_{n+1}}^{(0)} + \sum_{j=1}^{D} \delta_{t_{n+1}}^{(j)} \tilde{\Delta}^{(j)}, \nonumber
	\end{align}
  	where $\tilde{\Delta}^{(j)}$ is the $j^{\text{th}}$ option delta (i.e. $\tilde{\Delta}^{(j)}=\frac{\partial S_{t_{n}}^{(j,b)}}{\partial S_{t_{n}}^{(0,b)}}$). Note that $\tilde{\Delta}^{(j)}$ is time-independent since the calls and puts used for hedging are always of the same characteristics at each trading date (i.e. same moneyness and maturity) and both risky asset models are homoskedastic which entails that the underlying returns have the same conditional distribution for all time-steps. The $\tilde{\Delta}^{(j)}$ can be computed with the well-known closed form solutions under both models. For a total of $\tilde{N}$ simulated paths, the average exposure is computed as follows:
  	$$\bar{\Delta}^{(pf)}:= \frac{1}{\tilde{N}N} \sum_{k=1}^{\tilde{N}}\sum_{n=0}^{N-1}\tilde{\Delta}^{(pf)}_{t_{n},k},$$
  	where $\tilde{\Delta}^{(pf)}_{t_{n},k}$ is the time-$t_{n}$ portfolio delta of the $k^{\text{th}}$ simulated path. Results presented below for average exposures are from the test set.
  	
  	\subsubsection{Average exposure results}
  	\cref{table:equity_risk_exposure} presents average exposures of QDH and SQDH policies with the same market setup as in previous sections with respect to hedging instruments, model parameters and lookback option to hedge. The initial capital investments are again set as the risk-neutral price of the lookback option under each dynamics (i.e. $17.7\$$ and $25.3\$$ for BSM and MJD).
  	\begin{table}[!htbp]
	\caption {Average equity exposures with quadratic deep hedging (QDH) and semi-quadratic deep hedging (SQDH) for the lookback option of $T = 10$ years under the BSM and MJD model.} \label{table:equity_risk_exposure}
	\renewcommand{\arraystretch}{1.15}
		\begin{adjustwidth}{-1in}{-1in} 
	\centering
	\begin{tabular}{lccccc}
		\hline\noalign{\smallskip}
		& \multicolumn{2}{c}{$\text{BSM}$} & & \multicolumn{2}{c}{$\text{MJD}$} \\
		\cline{2-3} \cline{5-6}  & QDH  & SQDH & & QDH  & SQDH \\
		\hline\noalign{\medskip} 
		%
		Stock (year)  & $-0.10$ & $0.18$  & & $-0.14$  & $0.17$   \\
		Stock (month) & $-0.10$ & $-0.01$  & & $-0.15$  & $0.07$   \\
		Two options   & $-0.12$ & $-0.06$ & & $-0.10$  & $-0.04$  \\
		Six options   & $-0.12$ & $-0.11$ & & $-0.10$  & $-0.08$  \\
		\noalign{\medskip}\hline
	\end{tabular}%
	\end{adjustwidth}
	Notes: Average equity exposures under the BSM and MJD model with $S_{0}^{(0,b)} = 100$ and $r=0.03$. Both models dynamics under $\mathbb{P}$ and $\mathbb{Q}$ are described in \cref{subsec:dynamics_market} (see \cref{table:params_BSM} and \cref{table:params_MJD} for parameters values). Initial capital investments are respectively of $17.7\$$ and $25.3\$$ under BSM and MJD. \textit{Hedging instruments}: monthly and yearly underlying, yearly ATM call and put options (\textit{two options}) and three yearly calls and puts of strikes $K = \{S_{t_{n}}^{(0,b)}, 1.1S_{t_{n}}^{(0,b)}, 1.2S_{t_{n}}^{(0,b)}\}$ and $K = \{S_{t_{n}}^{(0,b)}, 0.9S_{t_{n}}^{(0,b)}, 0.8S_{t_{n}}^{(0,b)}\}$ (\textit{six options}). Results for QDH and SQDH are computed based on $75,\!000$ independent paths generated from the BSM and MJD model under $\mathbb{P}$. Training of the neural networks is done as described in \cref{subsubsec:hyperparameters}. 
	\end{table}
	Numerical results indicate that on average, SQDH policies are significantly more bullish than QDH policies under both dynamics and for all hedging instruments with a lesser extend for six options.
	This characteristic of SQDH policies to be more geared towards being long equity risk through a larger average exposure is most important with the underlying on a yearly basis, followed by monthly trading in the underlying, two options and six options. The observation that the average exposure of SQDH policies is only slightly larger than the average exposure of QDH policies when hedging with six options is consistent with benchmarks presented in previous sections.
	Indeed, values from \cref{table:BSM_MSE_vs_SMSE_no_TC} and \cref{table:MJD_MSE_vs_SMSE_no_TC} show that the absolute difference between the hedging statistics of QDH and SQDH is by far the smallest with six options. The latter naturally implies that the hedging positions of quadratic and non-quadratic policies are on average more similar with six options than with the other hedging instruments, which thus results in relatively closer average equity exposure.
	One direct implication of the larger average exposure of SQDH policies 
	is that in the risk management of the lookback option, SQDH 
	should result in positive expected gains. This was in fact observed in the benchmarking of global policies presented in \cref{table:BSM_MSE_vs_SMSE_no_TC} and \cref{table:MJD_MSE_vs_SMSE_no_TC} where SQDH resulted in negative mean hedging error statistics (i.e. mean hedging gains) under both risky assets dynamics.
	It is worth noting that \cite{trottier2018local} developed local risk minimization strategies for long-term options which also earned positive returns on average as well as reduced downside risk as compared to delta-hedging by having larger equity risk exposures.
	
	%

	\subsubsection{Analysis of SQDH bullishness}
	The distinctive feature of SQDH policies to hold a larger average equity exposure than with QDH can firstly be explained by the impact of hedging gains and losses on the optimized policies as measured by each penalty. On the one hand, by minimizing the MSE statistic in a market with positive expected log-returns for the underlying as implied by both models parameters values, QDH policies have to be less bullish whenever the hedging portfolio value at maturity is expected to be larger than the lookback option payoff. On the other hand, SQDH policies are strictly penalized for hedging losses proportionally to their squared values, not for hedging gains. The latter entails that SQDH policies are not constrained to reduce their equity risk exposure when the hedging portfolio value is expected to be larger than the lookback option payoff. The second important factor which contributes to SQDH bullishness specifically when hedging is done with the underlying is the capacity of deep agents to learn to benefit from \textit{time diversification of risk}. In the context of this study, time diversification of risk refers to the fact that investing in stocks over a long-term horizon reduces the risk of observing large losses as compared to short-term investments. Average exposure values in \cref{table:equity_risk_exposure} indicate that 
	deep agents hedging with the underlying and penalized with the SMSE have learned to hold a larger equity risk exposure than under the MSE penalty 
	to benefit simultaneously from the positive expected returns of the underlying and from the downside risk reduction with time diversification of risk. This observation is most important with the underlying on a yearly basis with SQDH obtaining average exposures of $0.18$ and $0.17$ under respectively the Black-Scholes and the MJD dynamics as compared to $-0.10$ and $-0.14$ with QDH.  
	
	Moreover, it is very interesting to note that the deep agents rely more on time diversification of risk in the presence of jump risk, i.e. with the MJD dynamics. Indeed, 
%
	the average exposure difference between SQDH and QDH policies with the underlying is significantly larger under the MJD dynamics with a difference of $0.31$ and $0.22$ for yearly and monthly trading as compared to $0.28$ and $0.09$ under the BSM.\footnote{
		For instance, the average exposure difference between SQDH and QDH with the underlying on a yearly basis under the MJD model is $0.17 - (-0.14) = 0.31$.
	} The latter observations can be explained by the fact that as shown in \cref{subsubsec:QDH_vs_SQDH}, hedging only with the underlying in the presence of jump risk is inefficient as compared to hedging with options. Thus, in the presence of jump risk, SQDH agents learn to rely more on time diversification of risk by having on average larger positions in the underlying as compared to SQDH agents trained on a Black-Scholes dynamics. 
	These findings thus provide additional evidence that the deep hedging algorithm is in fact model-free in the sense that the neural networks are able to effectively adapt their trading policies to different stylized facts of risky asset dynamics only by experiencing simulations of the financial market exhibiting these features.
	
	
	\section{Conclusion}
	\label{sec:conclusion}
	This paper studies global hedging strategies of long-term financial derivatives with a reinforcement learning approach. A similar financial market setup to the work of \cite{coleman2007robustly} is considered by studying the impact of equity risk with jump risk for the equity on the hedging effectiveness of segregated funds GMMBs. In the context of this paper, the latter guarantee is equivalent to holding a short position in a long-term lookback option of fixed maturity. The deep hedging algorithm of \cite{buehler2019deep} is applied to optimize long short-term memory networks representing global hedging policies
	with the mean-square error (MSE) and semi-mean-square error (SMSE) penalties 
	and with various hedging instruments (e.g. standard options and the underlying).
	
	
	Monte Carlo simulations are performed under the Black-Scholes model (BSM) and the Merton jump-diffusion (MJD) model to benchmark the hedging effectiveness of quadratic deep hedging (QDH) and semi-quadratic deep hedging (SQDH). Numerical results showed that under both dynamics and across all trading instruments, SQDH results in hedging policies which simultaneously reduce downside risk and increase expected returns as compared to QDH. The downside risk reduction achieved with SQDH over QDH ranges between $51\%$ to $85\%$ under the BSM and $45\%$ to $76\%$ under the MJD model. 
	Numerical experiments also indicated that QDH outperforms the local risk minimization scheme of \cite{coleman2007robustly} across all downside risk metrics and all hedging instruments. Thus, our results clearly demonstrate that SQDH policies should be prioritized as they are tailor-made to match the financial objectives of the hedger by significantly reducing downside risk as well as resulting in large expected positive returns.
	
	Monte Carlo experiments are also done to provide insight into specific characteristics of the optimized global policies. Numerical results showed that on average, SQDH policies 
	are significantly more bullish than QDH policies for every example considered. 
	Analysis presented in this paper indicate that the bullishness of SQDH policies stems from the impact of hedging gains and losses on the optimized policies as measured by each penalty.
	Furthermore, an additional factor which contributes to the larger average equity exposure of SQDH policies when hedging with the underlying is the capacity of deep agents to learn to benefit from time diversification of risk. The latter was shown to be most important in the presence of jump risk for the equity where deep agents penalized with the SMSE learned by experiencing many simulations of the financial market to rely more on time diversification risk through larger positions in the underlying as compared to training on the Black-Scholes dynamics due to the lesser efficiency of hedging with the underlying in the presence of jumps.
	
	Further research in the area of global hedging for long-term contingent claims with the deep hedging algorithm would prove worthwhile. The analysis of the impact of additional equity risk factors (e.g. volatility risk and regime risk) 
	on the optimized policies would be of interest. The same methodological approach presented in this paper could be applied with the addition of the latter equity risk factors with closed to no modification to the algorithm. Moreover, robustness analysis of the optimized policies when dynamics experienced slightly differ from the ones used to train the neural networks would prove worthwhile. The inclusion of realistic transaction costs for each hedging instrument could also be considered following the methodology of the original work of \cite{buehler2019deep}.


	\newpage
	\bibliographystyle{apalike}
	\bibliography{Biblio_paper_2}

\begin{thebibliography}{}

\bibitem[Abadi et~al., 2016]{abadi2016tensorflow}
Abadi, M. et~al. (2016).
\newblock Tensorflow: Large-scale machine learning on heterogeneous distributed
  systems.
\newblock {\em arXiv preprint arXiv:1603.04467}.

\bibitem[Almahdi and Yang, 2017]{almahdi2017adaptive}
Almahdi, S. and Yang, S.~Y. (2017).
\newblock An adaptive portfolio trading system: A risk-return portfolio
  optimization using recurrent reinforcement learning with expected maximum
  drawdown.
\newblock {\em Expert Systems with Applications}, 87:267--279.

\bibitem[Ankirchner et~al., 2014]{ankirchner2014cross}
Ankirchner, S., Schneider, J.~C., and Schweizer, N. (2014).
\newblock Cross-hedging minimum return guarantees: Basis and liquidity risks.
\newblock {\em Journal of Economic Dynamics and Control}, 41:93--109.

\bibitem[Augustyniak and Boudreault, 2017]{augustyniak2017mitigating}
Augustyniak, M. and Boudreault, M. (2017).
\newblock Mitigating interest rate risk in variable annuities: An analysis of
  hedging effectiveness under model risk.
\newblock {\em North American Actuarial Journal}, 21(4):502--525.

\bibitem[Augustyniak et~al., 2017]{augustyniak2017assessing}
Augustyniak, M., Godin, F., and Simard, C. (2017).
\newblock Assessing the effectiveness of local and global quadratic hedging
  under \text{GARCH} models.
\newblock {\em Quantitative Finance}, 17(9):1305--1318.

\bibitem[Bacinello, 2003]{bacinello2003fair}
Bacinello, A.~R. (2003).
\newblock Fair valuation of a guaranteed life insurance participating contract
  embedding a surrender option.
\newblock {\em Journal of risk and insurance}, 70(3):461--487.

\bibitem[Bauer et~al., 2008]{bauer2008universal}
Bauer, D., Kling, A., and Russ, J. (2008).
\newblock A universal pricing framework for guaranteed minimum benefits in
  variable annuities.
\newblock {\em ASTIN Bulletin: The Journal of the IAA}, 38(2):621--651.

\bibitem[Becker et~al., 2019]{becker2019deep}
Becker, S., Cheridito, P., and Jentzen, A. (2019).
\newblock Deep optimal stopping.
\newblock {\em Journal of Machine Learning Research}, 20:1--25.

\bibitem[Bengio et~al., 1994]{bengio1994learning}
Bengio, Y., Simard, P., and Frasconi, P. (1994).
\newblock Learning long-term dependencies with gradient descent is difficult.
\newblock {\em IEEE transactions on neural networks}, 5(2):157--166.

\bibitem[Bertsimas et~al., 2001]{bertsimas2001hedging}
Bertsimas, D., Kogan, L., and Lo, A.~W. (2001).
\newblock Hedging derivative securities and incomplete markets: an
  $\epsilon$-arbitrage approach.
\newblock {\em Operations Research}, 49(3):372--397.

\bibitem[Black and Scholes, 1973]{black1973pricing}
Black, F. and Scholes, M. (1973).
\newblock The pricing of options and corporate liabilities.
\newblock {\em Journal of Political Economy}, 81(3):637--654.

\bibitem[Boyle and Hardy, 1997]{boyle1997reserving}
Boyle, P.~P. and Hardy, M.~R. (1997).
\newblock Reserving for maturity guarantees: Two approaches.
\newblock {\em Insurance: Mathematics and Economics}, 21(2):113--127.

\bibitem[Boyle and Schwartz, 1977]{boyle1977equilibrium}
Boyle, P.~P. and Schwartz, E.~S. (1977).
\newblock Equilibrium prices of guarantees under equity-linked contracts.
\newblock {\em Journal of Risk and Insurance}, 44:639--660.

\bibitem[Brennan and Schwartz, 1976]{brennan1976pricing}
Brennan, M.~J. and Schwartz, E.~S. (1976).
\newblock The pricing of equity-linked life insurance policies with an asset
  value guarantee.
\newblock {\em Journal of Financial Economics}, 3(3):195--213.

\bibitem[Buehler et~al., 2019a]{buehler2019deep}
Buehler, H., Gonon, L., Teichmann, J., and Wood, B. (2019a).
\newblock Deep hedging.
\newblock {\em Quantitative Finance}, 19(8):1271--1291.

\bibitem[Buehler et~al., 2019b]{buehler2019deep_2}
Buehler, H., Gonon, L., Teichmann, J., Wood, B., Mohan, B., and Kochems, J.
  (2019b).
\newblock Deep hedging: hedging derivatives under generic market frictions
  using reinforcement learning.
\newblock Technical Report 19-80.

\bibitem[Carbonneau and Godin, 2020]{carbonneau2020}
Carbonneau, A. and Godin, F. (2020).
\newblock Equal risk pricing of derivatives with deep hedging.
\newblock {\em arXiv preprint arXiv:2002.08492}.

\bibitem[Coleman et~al., 2007]{coleman2007robustly}
Coleman, T., Kim, Y., Li, Y., and Patron, M. (2007).
\newblock Robustly hedging variable annuities with guarantees under jump and
  volatility risks.
\newblock {\em Journal of Risk and Insurance}, 74(2):347--376.

\bibitem[Coleman et~al., 2006]{coleman2006hedging}
Coleman, T., Li, Y., and Patron, M. (2006).
\newblock Hedging guarantees in variable annuities under both equity and
  interest rate risks.
\newblock {\em Insurance: Mathematics and Economics}, 38(2):215--228.

\bibitem[Delbaen and Schachermayer, 1994]{delbaen1994general}
Delbaen, F. and Schachermayer, W. (1994).
\newblock A general version of the fundamental theorem of asset pricing.
\newblock {\em Mathematische Annalen}, 300(1):463--520.

\bibitem[Deng et~al., 2016]{deng2016deep}
Deng, Y. et~al. (2016).
\newblock Deep direct reinforcement learning for financial signal
  representation and trading.
\newblock {\em IEEE Transactions on Neural Networks and Learning Systems},
  28(3):653--664.

\bibitem[Dupuis et~al., 2016]{dupuis2016short}
Dupuis, D., Gauthier, G., and Godin, F. (2016).
\newblock Short-term hedging for an electricity retailer.
\newblock {\em The Energy Journal}, 37(2):31--59.

\bibitem[F{\"o}llmer and Schweizer, 1988]{follmer1988hedging}
F{\"o}llmer, H. and Schweizer, M. (1988).
\newblock Hedging by sequential regression: An introduction to the mathematics
  of option trading.
\newblock {\em ASTIN Bulletin: The Journal of the IAA}, 18(2):147--160.

\bibitem[Fran{\c{c}}ois et~al., 2014]{franccois2014optimal}
Fran{\c{c}}ois, P., Gauthier, G., and Godin, F. (2014).
\newblock Optimal hedging when the underlying asset follows a regime-switching
  markov process.
\newblock {\em European Journal of Operational Research}, 237(1):312--322.

\bibitem[Gan, 2013]{gan2013application}
Gan, G. (2013).
\newblock Application of data clustering and machine learning in variable
  annuity valuation.
\newblock {\em Insurance: Mathematics and Economics}, 53(3):795--801.

\bibitem[Glorot and Bengio, 2010]{glorot2010understanding}
Glorot, X. and Bengio, Y. (2010).
\newblock Understanding the difficulty of training deep feedforward neural
  networks.
\newblock In {\em Proceedings of the thirteenth international conference on
  artificial intelligence and statistics}, pages 249--256.

\bibitem[Godin, 2016]{godin2016minimizing}
Godin, F. (2016).
\newblock Minimizing \text{CVaR} in global dynamic hedging with transaction
  costs.
\newblock {\em Quantitative Finance}, 16(3):461--475.

\bibitem[Goodfellow et~al., 2016]{goodfellow2016deep}
Goodfellow, I., Bengio, Y., and Courville, A. (2016).
\newblock {\em Deep learning}.
\newblock MIT press.

\bibitem[Halperin, 2020]{halperin2020qlbs}
Halperin, I. (2020).
\newblock Qlbs: Q-learner in the black-scholes (-merton) worlds.
\newblock {\em The Journal of Derivatives}.

\bibitem[Han and E, 2016]{han2016deep}
Han, J. and E, W. (2016).
\newblock Deep learning approximation for stochastic control problems.
\newblock {\em arXiv preprint arXiv:1611.07422}.

\bibitem[Hardy, 2003]{hardy2003investment}
Hardy, M. (2003).
\newblock {\em Investment guarantees: modeling and risk management for
  equity-linked life insurance}, volume 215.
\newblock John Wiley \& Sons.

\bibitem[Hardy, 2000]{hardy2000hedging}
Hardy, M.~R. (2000).
\newblock Hedging and reserving for single-premium segregated fund contracts.
\newblock {\em North American Actuarial Journal}, 4(2):63--74.

\bibitem[Harrison and Pliska, 1981]{harrison1981martingales}
Harrison, J.~M. and Pliska, S.~R. (1981).
\newblock Martingales and stochastic integrals in the theory of continuous
  trading.
\newblock {\em Stochastic Processes and their Applications}, 11(3):215--260.

\bibitem[Hochreiter and Schmidhuber, 1997]{hochreiter1997long}
Hochreiter, S. and Schmidhuber, J. (1997).
\newblock Long short-term memory.
\newblock {\em Neural computation}, 9(8):1735--1780.

\bibitem[Hongkai et~al., 2020]{hongkai2020}
Hongkai, C., Cui, Z., and Yanchu, L. (2020).
\newblock Discrete-time variance-optimal deep hedging in affine \text{GARCH}
  models.
\newblock {\em Working paper}.

\bibitem[Jiang et~al., 2017]{jiang2017deep}
Jiang, Z., Xu, D., and Liang, J. (2017).
\newblock A deep reinforcement learning framework for the financial portfolio
  management problem.
\newblock {\em arXiv preprint arXiv:1706.10059}.

\bibitem[K{\'e}lani and Quittard-Pinon, 2017]{kelani2017pricing}
K{\'e}lani, A. and Quittard-Pinon, F. (2017).
\newblock Pricing and hedging variable annuities in a \text{L{\'e}vy} market: a
  risk management perspective.
\newblock {\em Journal of Risk and Insurance}, 84(1):209--238.

\bibitem[Kingma and Ba, 2014]{kingma2014adam}
Kingma, D.~P. and Ba, J. (2014).
\newblock Adam: A method for stochastic optimization.
\newblock {\em arXiv preprint arXiv:1412.6980}.

\bibitem[Kolm and Ritter, 2019]{kolm2019dynamic}
Kolm, P.~N. and Ritter, G. (2019).
\newblock Dynamic replication and hedging: A reinforcement learning approach.
\newblock {\em The Journal of Financial Data Science}, 1(1):159--171.

\bibitem[Lamberton and Lapeyre, 2011]{lamberton2011introduction}
Lamberton, D. and Lapeyre, B. (2011).
\newblock {\em Introduction to stochastic calculus applied to finance}.
\newblock Chapman and Hall/CRC.

\bibitem[Li et~al., 2009]{li2009learning}
Li, Y., Szepesvari, C., and Schuurmans, D. (2009).
\newblock Learning exercise policies for american options.
\newblock In {\em Artificial Intelligence and Statistics}, pages 352--359.

\bibitem[Merton, 1976]{Merton1976}
Merton, R.~C. (1976).
\newblock Option pricing when underlying stock returns are discontinuous.
\newblock {\em Journal of Financial Economics}, 3:125--144.

\bibitem[Moody and Saffell, 2001]{moody2001learning}
Moody, J. and Saffell, M. (2001).
\newblock Learning to trade via direct reinforcement.
\newblock {\em IEEE Transactions on Neural Networks}, 12(4):875--889.

\bibitem[Persson and Aase, 1997]{persson1997valuation}
Persson, S.-A. and Aase, K.~K. (1997).
\newblock Valuation of the minimum guaranteed return embedded in life insurance
  products.
\newblock {\em Journal of Risk and Insurance}, 64(4):599--617.

\bibitem[Powell, 2009]{powell2009you}
Powell, W.~B. (2009).
\newblock What you should know about approximate dynamic programming.
\newblock {\em Naval Research Logistics (NRL)}, 56(3):239--249.

\bibitem[R{\'e}millard and Rubenthaler, 2013]{remillard2013optimal}
R{\'e}millard, B. and Rubenthaler, S. (2013).
\newblock Optimal hedging in discrete time.
\newblock {\em Quantitative Finance}, 13(6):819--825.

\bibitem[Rockafellar and Uryasev, 2002]{rockafellar2002conditional}
Rockafellar, R.~T. and Uryasev, S. (2002).
\newblock Conditional \text{Value-at-Risk} for general loss distributions.
\newblock {\em Journal of Banking \& Finance}, 26(7):1443--1471.

\bibitem[Rumelhart et~al., 1986]{rumelhart1986learning}
Rumelhart, D.~E., Hinton, G.~E., and Williams, R.~J. (1986).
\newblock Learning representations by back-propagating errors.
\newblock {\em Nature}, 323(6088):533--536.

\bibitem[Schweizer, 1991]{schweizer1991option}
Schweizer, M. (1991).
\newblock Option hedging for semimartingales.
\newblock {\em Stochastic processes and their Applications}, 37(2):339--363.

\bibitem[Schweizer, 1995]{schweizer1995variance}
Schweizer, M. (1995).
\newblock Variance-optimal hedging in discrete time.
\newblock {\em Mathematics of Operations Research}, 20(1):1--32.

\bibitem[Trottier et~al., 2018a]{trottier2018local}
Trottier, D.-A., Godin, F., and Hamel, E. (2018a).
\newblock Local hedging of variable annuities in the presence of basis risk.
\newblock {\em ASTIN Bulletin: The Journal of the IAA}, 48(2):611--646.

\bibitem[Trottier et~al., 2018b]{trottier2018fund}
Trottier, D.-A., Godin, F., and Hamel, E. (2018b).
\newblock On fund mapping regressions applied to segregated funds hedging under
  regime-switching dynamics.
\newblock {\em Risks}, 6(3):78.

\bibitem[Zhang, 2010]{zhang2010integrating}
Zhang, F. (2010).
\newblock Integrating robust risk management into pricing: New thinking for
  \text{VA} writers.
\newblock {\em Risk and Rewards}, 55:34--36.

\end{thebibliography}
	
	\end{document}